\documentclass[sigconf, nonacm, screen, balance]{acmart}
\newcommand\vldbdoi{10.14778/3801059.3801064}
\newcommand\vldbpages{XXX-XXX}
\newcommand\vldbvolume{19}
\newcommand\vldbissue{7}
\newcommand\vldbyear{2026}
\newcommand\vldbauthors{\authors}
\newcommand\vldbtitle{\shorttitle} 
\newcommand\vldbavailabilityurl{https://github.com/LeonLee666/IObench4DiskANN}
\newcommand\vldbpagestyle{plain} 

\usepackage[]{mypkg}
\usepackage{balance}
\usepackage[normalem]{ulem}
\usepackage{todonotes}
\usepackage{marginnote}
\newcommand{\stitle}[1]{\vspace{1.2ex}\noindent{\bf #1}}

\newcommand{\eat}[1]{}

\begin{document}

\title{I/O Optimizations for Graph-Based Disk-Resident Approximate Nearest Neighbor Search: A Design Space Exploration}

\author{Liang~Li}
\orcid{0000-0002-2527-5049}
\affiliation{%
  \institution{China Telecom Cloud Computing Research Institute}
  \postcode{43017-6221}
}
\email{lil225@chinatelecom.cn}

\author{Shufeng~Gong}
\affiliation{%
  \institution{Northeastern University}
}
\email{gongsf@mail.neu.edu.cn}

\author{Yanan Yang}
\authornote{Corresponding author: Yanan~Yang.}
\affiliation{%
  \institution{China Telecom Cloud Computing Research Institute}
  \postcode{43017-6221}
}
\email{yangyn11@chinatelecom.cn}

\author{Yiduo~Wang}
\affiliation{%
  \institution{China Telecom Cloud Computing Research Institute}
  \postcode{43017-6221}
}
\email{wangyd22@chinatelecom.cn}

\author{Jie~Wu}
\affiliation{%
  \institution{China Telecom Cloud Computing Research Institute}
}
\affiliation{%
  \institution{Temple University}
}
\email{jiewu@temple.edu}

\begin{abstract}
    Approximate nearest neighbor (ANN) search on SSD-backed indexes is increasingly I/O-bound (I/O accounts for 70--90\% of query latency). We present an I/O-first framework for disk-based ANN that organizes techniques along three dimensions: memory layout, disk layout, and search algorithm. We introduce a page-level complexity model that explains how page locality and path length jointly determine page reads, and we validate the model empirically. Using consistent implementations across four public datasets, we quantify both single-factor effects and cross-dimensional synergies. We find that (i) memory-resident navigation and dynamic width provide the strongest standalone gains; (ii) page shuffle and page search are weak alone but complementary together; and (iii) a principled composition, OctopusANN, substantially reduces I/O and achieves 4.1--37.9\% higher throughput than the state-of-the-art system Starling and 87.5--149.5\% higher throughput than DiskANN at matched Recall@10=90\%. Finally, we distill actionable guidelines for selecting storage-centric or hybrid designs across diverse concurrency levels and accuracy constraints, advocating systematic composition rather than isolated tweaks when pushing the performance frontier of disk-based ANN.
\end{abstract}
\maketitle

\pagestyle{\vldbpagestyle}
\begingroup\small\noindent\raggedright\textbf{PVLDB Reference Format:}\\
\vldbauthors. \vldbtitle. PVLDB, \vldbvolume(\vldbissue): \vldbpages, \vldbyear.\\
\href{https://doi.org/\vldbdoi}{doi:\vldbdoi}
\endgroup
\begingroup
\renewcommand\thefootnote{}\footnote{\noindent
This work is licensed under the Creative Commons BY-NC-ND 4.0 International License. Visit \url{https://creativecommons.org/licenses/by-nc-nd/4.0/} to view a copy of this license. For any use beyond those covered by this license, obtain permission by emailing \href{mailto:info@vldb.org}{info@vldb.org}. Copyright is held by the owner/author(s). Publication rights licensed to the VLDB Endowment. \\
\raggedright Proceedings of the VLDB Endowment, Vol. \vldbvolume, No. \vldbissue\ %
ISSN 2150-8097. \\
\href{https://doi.org/\vldbdoi}{doi:\vldbdoi} \\
}\addtocounter{footnote}{-1}\endgroup

\ifdefempty{\vldbavailabilityurl}{}{
\vspace{.3cm}
\begingroup\small\noindent\raggedright\textbf{PVLDB Artifact Availability:}\\
The source code, data, and/or other artifacts have been made available at \url{\vldbavailabilityurl}.
\endgroup
}

\section{Introduction}
\label{sec:introduction}
Approximate nearest neighbor (ANN) search is fundamental to modern retrieval-intensive applications~\cite{Aoyama2011,Arora2018,Fu2017,Malkov2020,Wang2021a}, including information retrieval~\cite{Flickner2002,Jiang2019}, pattern recognition~\cite{Cover,Kosuge2019}, data mining~\cite{Huang2017,Iwasaki2016}, machine learning~\cite{Cao2018,Cost1993}, recommendation systems~\cite{Meng2019,Sarwar2001}, retrieval-augmented generation (RAG)~\cite{Lewis2020}, and vector databases~\cite{Wang2021a,Guo2022,Wei2020}. Among various approaches—hashing-based~\cite{Gong2020,Huang2015}, tree-based~\cite{Arora2018,SilpaAnan2008}, and quantization-based~\cite{Jegou2011,Pan2020,Gao2024}—graph-based algorithms have emerged as state of the art for search quality and efficiency~\cite{Wang2021,Azizi2025,Yang2024}. Representative methods such as HNSW~\cite{Malkov2020}, NSG~\cite{Fu2017}, SSG~\cite{Fu2022}, and Vamana~\cite{Subramanya2019} achieve exceptional performance with high recall and low latency. However, their substantial memory overhead renders them infeasible for billion-scale datasets on commodity hardware~\cite{Subramanya2019,Chen2021,Wang2024}.

As datasets scale to billions of vectors, the prohibitive cost of DRAM necessitates a shift toward disk-resident solutions for vector search~\cite{Subramanya2019,simhadri2021freshdiskann,Gollapudi2023}. While in-memory indices have been extensively studied~\cite{Wang2021,Azizi2025,Yang2024,Li_Zhang_Sun_Wang_Li_Zhang_Lin_2016}, disk-resident graph indices face fundamentally different challenges driven by the I/O bottleneck. 
Unlike in-memory search, which is compute-bound, disk-based graph traversal becomes I/O-bound due to \textbf{high I/O latency} and \textbf{random I/O access patterns}: each hop may trigger fine-grained random reads whose per-request latency dominates distance computation. 
This fundamental shift leads to severe performance degradation and an optimization landscape that differs markedly from that of in-memory systems. Consequently, this work focuses exclusively on improving the I/O efficiency of disk-based graph indices.

Among existing disk-based solutions, DiskANN~\cite{Subramanya2019} has established itself as the \textit{de facto} industrial standard~\cite{Simhadri2022,Simhadri2024}, widely adopted by major commercial vector databases including Microsoft Azure~\cite{Microsoft2025} and Huawei GaussDB~\cite{HuaweiCloud2024}, \etc. It employs a three-layer architecture: a disk storage layer for the full graph and vectors, a memory layer for compressed navigation vectors, and a cache for high-degree nodes. Despite its robust design, our analysis reveals that I/O operations still account for 70--90\% of total query latency. This bottleneck stems from four critical inefficiencies: (1) \emph{poor data locality}, where scattered vertices cause massive I/O waste~\cite{Wang2024}; (2) \emph{low bandwidth utilization} due to strictly sequential I/O-compute execution~\cite{async-diskann,guoachieving}; (3) \emph{long search paths} that require excessive disk round-trips~\cite{Subramanya2019,Wang2024,guoachieving}; and (4) \emph{frequent cache misses} resulting from unpredictable random access patterns~\cite{async-diskann,Subramanya2019,Jang2023,Jang2024}.

There has been extensive work addressing these issues, including caching strategies~\cite{async-diskann,Subramanya2019,Jang2023}, quantization methods~\cite{Zhang2019,Subramanya2019}, hierarchical structures~\cite{Wang2024,Zhang2019}, and I/O pipelining approaches~\cite{guoachieving}.
Although several studies~\cite{Wang2021,Azizi2025,Yang2024,Li_Zhang_Sun_Wang_Li_Zhang_Lin_2016} have examined graph-based indexes, no survey systematically explores I/O optimizations. This gap motivates us to conduct an in-depth investigation of I/O efficiency. This paper explores I/O optimization methods for on-disk graph-based ANN indexes and focuses on answering the following three questions.

\stitle{Q1: How to classify I/O efficiency optimizations systematically to highlight the differences and connections among various methods?} 
Existing optimization techniques are studied in isolation without a unified classification framework. Although numerous approaches have been proposed—vector quantization~\cite{Zhang2019,Subramanya2019}, hierarchical structures~\cite{Wang2024,Chen2021}, I/O pipelining~\cite{guoachieving,async-diskann}, and reorganization of data layout~\cite{Wang2024,Tatsuno2025}—they target different I/O bottlenecks without a systematic understanding of their orthogonal boundaries, making it difficult to navigate the design space.

\stitle{Q2:  What is the effectiveness of each individual optimization and its contribution to I/O performance?} Some optimization methods, such as pipeline and dynamic width~\cite{guoachieving}, have been shown to work synergistically, significantly improving I/O performance. However, the effectiveness of each of these optimizations when applied independently remains unclear.

\stitle{Q3: Do combinations of individual optimizations yield multiplicative performance improvements?} While state-of-the-art solutions such as Starling~\cite{Wang2024} and PipeANN~\cite{guoachieving} represent significant advances, systematic investigation into whether multidimensional optimization combinations can achieve breakthrough performance improvements remains lacking. The potential for carefully orchestrated combinations to deliver synergistic performance gains is largely unexplored.

To address these questions, this paper presents a systematic study of the I/O optimization design space for disk-based ANN search algorithms. We introduce a comprehensive three-dimensional taxonomy that organizes optimization techniques into orthogonal categories: \textit{(\romannumeral1) memory layout optimization}, \textit{(\romannumeral2) disk layout optimization}, and \textit{(\romannumeral3) search algorithm optimization}. Using this framework and consistent experimental methodology across four public datasets, we conduct controlled evaluations that quantify both individual optimization impacts and reveal synergistic combinations. Our key insight is that carefully orchestrated multi-dimensional approaches yield synergistic improvements by jointly increasing page-level utility and shortening convergence paths.

Our work makes three key contributions:

\stitle{(1) Systematic taxonomy framework.} We establish a comprehensive taxonomy for I/O optimization algorithms, identifying three orthogonal dimensions that unify previously fragmented approaches and provide an objective function for the navigation optimization landscape.

\stitle{(2)
Optimization evaluation and empirical findings.} We conduct a controlled, apples-to-apples evaluation across four massive datasets to quantify the effectiveness and costs of individual I/O optimizations, and to extract a set of empirical findings that characterize when each technique is beneficial, marginal, or counterproductive.

\stitle{(3) Combination case study and actionable guidelines.} Building on the above findings, we further study how orthogonal optimizations can be composed in practice, and present \emph{OctopusANN} as a representative effective combination. It substantially improves upon existing methods—achieving 4.1--37.9\% higher throughput than Starling and 87.5--149.5\% over DiskANN at 90\% recall—while our findings provide evidence-based guidelines for selecting and combining techniques under different workload and resource regimes.

The remainder of this paper is organized as follows. \S\ref{sec:background} provides essential background on graph-based ANN search, with detailed analysis of DiskANN's core design. \S\ref{sec:challenges} presents our comprehensive analysis of I/O challenges. \S\ref{sec:design_space} introduces our three-dimensional taxonomy and systematic analysis of optimization techniques. \S\ref{sec:experiments} describes our experimental testbed and methodology, while \S\ref{sec:exp-1} and \S\ref{sec:exp-2} report results for individual optimizations and their synergistic combinations, respectively. Finally, \S\ref{sec:conclusion} concludes the paper.

\section{Background}
\label{sec:background}

This section provides essential background on approximate nearest neighbor search, introducing fundamental concepts and challenges, followed by an analysis of graph-based ANN algorithms with emphasis on disk-based approaches.

\subsection{Nearest Neighbor Search}
\textbf{Approximate Nearest Neighbor Search.}
Given a dataset $D \subset \mathbb{R}^d$ containing $n$ points, the $k$-nearest neighbor ($k$-NN) search problem seeks the $k$ closest points $S^\ast$ to query point $q \in \mathbb{R}^d$, i.e., $p\in D\setminus S^\ast$, $p'\in S^\ast$, $||p,q|| \geq ||p',q||$, where $||\cdot||$ is a distance function, e.g., Euclidean or Cosine distance. While exact and robust search requires $O(nd)$ time complexity~\cite{Li_Zhang_Sun_Wang_Li_Zhang_Lin_2016}, approximate nearest neighbor search (ANN) trades accuracy for efficiency by returning approximate near neighbors $S$ at lower computational cost. The query accuracy of ANN methods is measured by the \textit{recall} value, $\mathrm{Recall@}k = \frac{|S \cap S^\ast|}{k}$. 

\stitle{Graph-Based ANN methods.}
Among various ANN search approaches, proximity-graph methods~\cite{Mathieson2019,Riegger2010} are widely used. Prior studies~\cite{Aumüller_Bernhardsson_Faithfull_2020,Li_Zhang_Andersen_He_2020,Li_Zhang_Sun_Wang_Li_Zhang_Lin_2016,Wang2021} report strong performance for graph-based methods. Graph-based ANN search constructs graphs where vertices represent vectors and edges connect similar vectors. Main approaches include Delaunay graphs~\cite{FORTUNE1992} (good geometric properties), $k$-nearest neighbor graphs~\cite{Paredes2005} (predictable degree), relative neighborhood graphs~\cite{Toussaint1980} (sparse connectivity), minimum spanning trees~\cite{Kruskal2010} (minimal edges), and small-world graphs~\cite{Malkov2014} (efficient for search). Modern algorithms such as HNSW~\cite{Malkov2020}, NSG~\cite{Fu2017}, SSG~\cite{Fu2022}, NGT~\cite{NGT}, SPTAG~\cite{SPTAG}, Vamana~\cite{Subramanya2019}, $\tau$-MG~\cite{Peng2023}, and ELPIS~\cite{Azizi2023} have shown strong performance in practice.

\stitle{Best-first search.} Graph-based ANN search commonly employs best-first search~\cite{Hart1968a} or beam search~\cite{Tellez2017}, maintaining a priority queue of candidate nodes sorted by distance to the query. The search iteratively selects the closest unvisited node (beam search is a width-limited variant in which each iteration selects the closest top-$\mathcal{\omega}$ unvisited nodes, where $\mathcal{\omega}$ is the \textit{beam width}), explores neighbors, and adds new candidates until identifying the required nearest neighbors. Search efficiency depends on entry point quality~\cite{Wang2021,Azizi2025}, graph structure quality~\cite{Boutet2016,Yang2024}, distance computation~\cite{Deng2024,Chen2023,Gao2023,Wei2025,Andre2017}, and early termination strategies~\cite{Li2020}.

\stitle{Limitations of graph-based ANN methods.} While in-memory graph-based methods achieve excellent search performance, they face significant memory limitations with large-scale (\eg billion-scale) datasets. Memory overhead increases substantially with dataset size—for example, HNSW on a billion-scale dataset in 128 dimensions can consume hundreds of gigabytes of memory (\eg ~800~GB under common settings), often exceeding typical workstation RAM capacity. When dataset sizes exceed available memory, traditional in-memory approaches become infeasible, prompting disk-based ANN search algorithms.

\begin{figure}[t]
    \centering
    \includegraphics[width=0.9\linewidth]{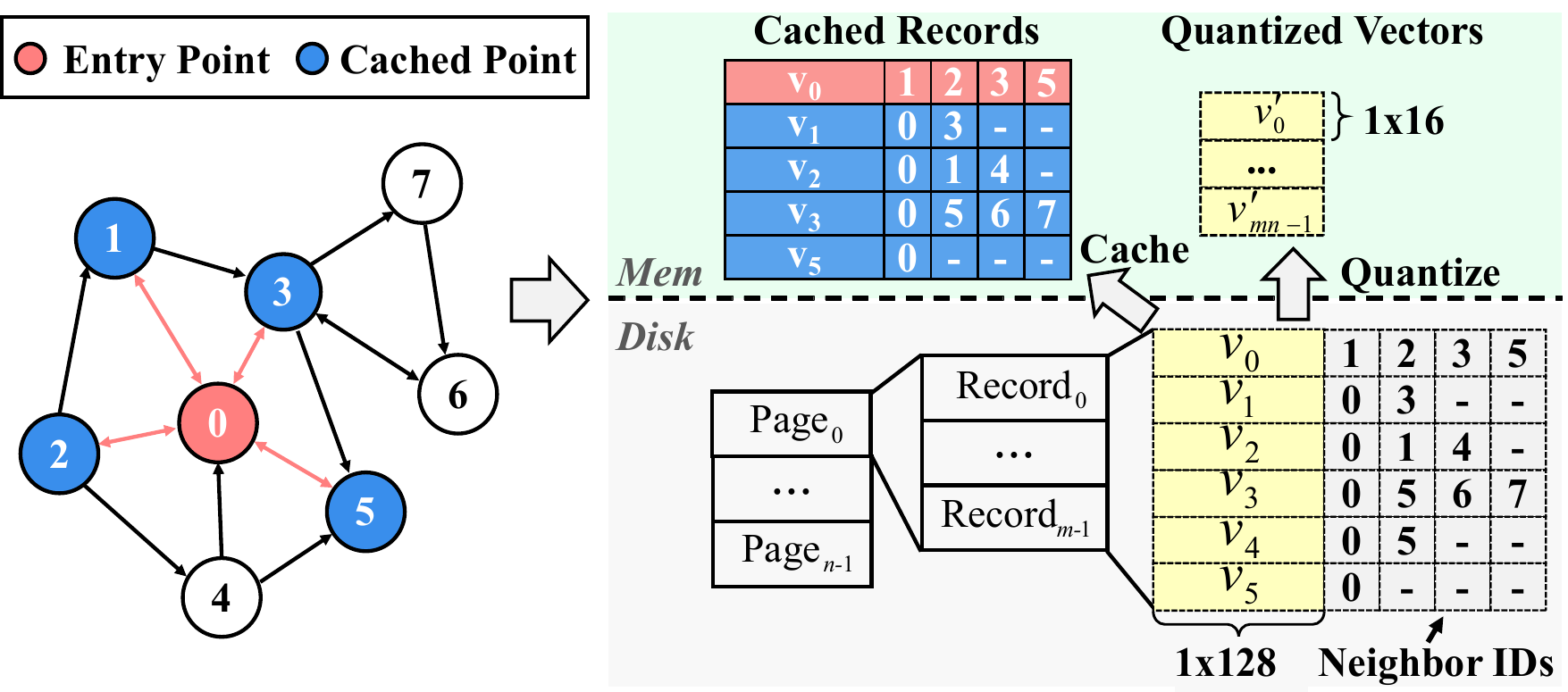}
    \Description{DiskANN disk layout diagram.}
    \mycaption{0}{-17}{DiskANN data layout. Each record is page-aligned and packed into 4~KB pages.}
    \label{fig:diskann-record}
\end{figure}

\subsection{Disk-Based ANN}
\label{sec:diskbased_overview}

To address the memory constraints that necessitate disk-resident search, we adopt DiskANN as a canonical model to illustrate how a three-layer architecture organizes on-disk vectors and graph traversal, setting the stage for our I/O-centric analysis in \S\ref{sec:challenges}.

\stitle{DiskANN overview and architecture.}
DiskANN~\cite{Subramanya2019,simhadri2021freshdiskann,Gollapudi2023,Jaiswal2022} is a \textit{de facto} standard disk-based ANN system adopted by products such as Milvus~\cite{Zilliz2025}, Azure Database~\cite{Microsoft2025}, Timescale~\cite{Timescale2025}, and GaussDB~\cite{HuaweiCloud2024}, and serves as a standard baseline in public benchmarks~\cite{Simhadri2022,Simhadri2024}. As shown in Figure~\ref{fig:diskann-record}, DiskANN follows a three-layer architecture: (1) a \textbf{disk storage layer} that stores full-precision vectors and neighbor lists in page-aligned records; (2) a \textbf{memory quantization layer} (\eg Product Quantization, \textbf{PQ}) that keeps compressed codes for fast, memory-resident filtering; and (3) a \textbf{cache management layer} that retains hot vertices. This design implements an ``approximate guidance, precise refinement'' workflow: memory-resident PQ codes guide candidate selection, while disk-resident full vectors enable final re-ranking.

\stitle{Representative disk-based systems.}
Beyond DiskANN, representative systems can be categorized into three orthogonal routes. \textit{Starling/DiskANN++}~\cite{Wang2024,Ni2023} are layout-centered, employing an in-memory navigation graph and locality-aware page shuffling (optionally in-page search) to raise per-page utility and shorten search paths. \textit{PipeANN}~\cite{guoachieving} is algorithm-centric, using I/O--compute pipelining with dynamic width; it may trigger speculative reads under concurrency. \textit{AiSAQ/LM-DiskANN}~\cite{Tatsuno2025,Pan2023} are storage-centric, placing PQ codes on SSD to minimize memory usage but risking read amplification and a larger on-disk footprint. Other variants include \textit{GRIP/Zoom}~\cite{Zhang2019,Zhang2018} (memory-first: PQ with hierarchical navigation/MemGraph under moderate memory budgets) and \textit{SPANN/BBANN}~\cite{Chen2021,BBANN} (partitioned/hierarchical navigators with storage-aware locality). This organization directly impacts per-page utility and traversal hops, which we analyze through an I/O-centric lens in \S\ref{sec:challenges}. We use DiskANN as a running reference to anchor later analysis of I/O efficiency challenges and to provide a consistent baseline for design-space comparisons.

\subsection{I/O is the Main Bottleneck}
\begin{figure}[t]
    \centering
    \includegraphics[width=0.9\linewidth]{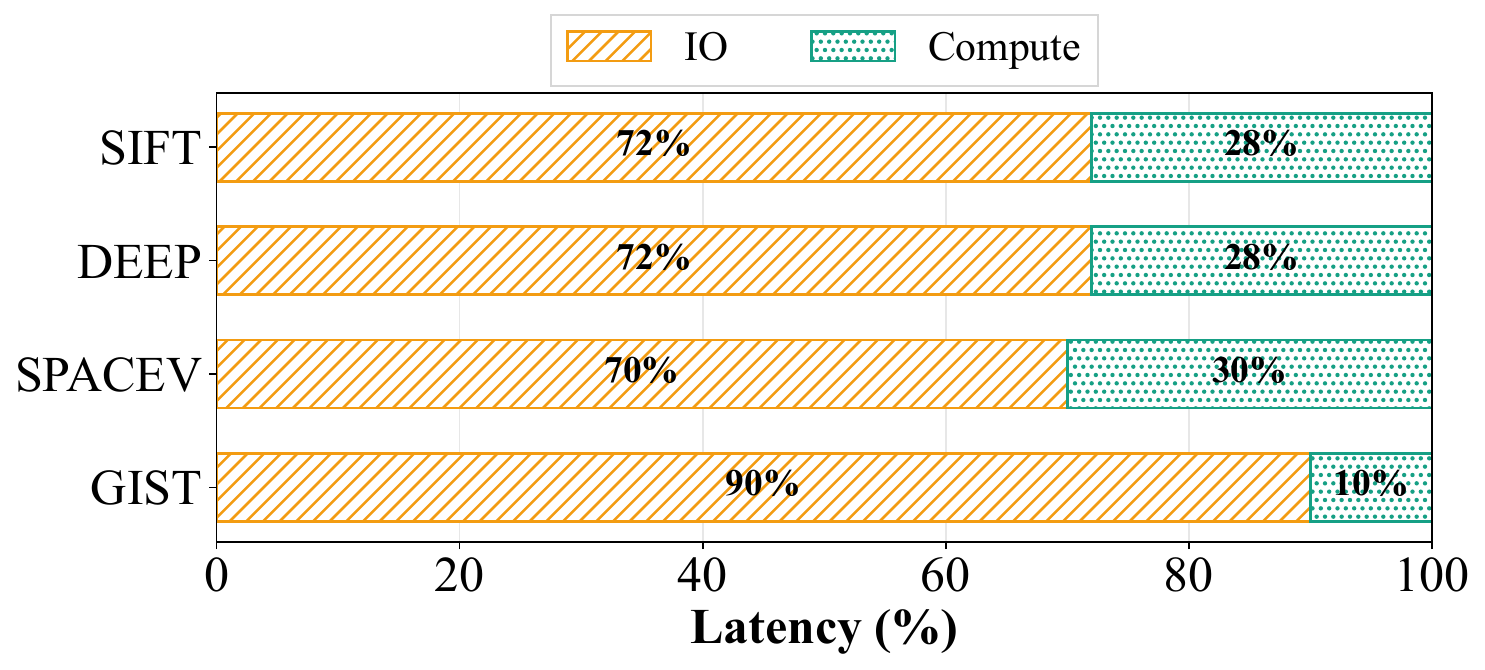}
    \Description{Latency vs recall comparison.}
    \mycaption{0}{-17}{Latency breakdown across four datasets.}
    \label{fig:io_profiling}   
\end{figure}

Although disk-based graph indices address the memory bottleneck of in-memory indices, they still face significant I/O efficiency challenges~\cite{Wang2024,guoachieving}. Figure~\ref{fig:io_profiling} shows the percentage breakdown of compute and I/O latency for DiskANN queries across four datasets. The figure reveals that I/O latency dominates query latency, accounting for approximately 70\%--90\% of the total (note that dataset descriptions are provided in \S\ref{sec:experiments}; the GIST dataset has higher dimensionality and requires more search hops). This dominance stems from the fundamental nature of graph-based vector search: 
(1) \textbf{High I/O latency}, where fetching nodes from disk is orders of magnitude slower than in-memory distance computation;
(2) \textbf{Random I/O access patterns}, as graph traversal inherently induces fine-grained random I/Os that fail to exploit the internal parallelism and throughput of modern SSDs.

\section{I/O Challenges Analysis}
\label{sec:challenges}

\subsection{I/O Complexity for Best-First Search}

To establish a formal foundation for analyzing I/O efficiency, we first introduce key terminology specific to graph-structured disk-resident indices, then present the I/O complexity model.

\stitle{Overlap Ratio.} To ground the I/O complexity analysis, we first formalize page-level utility via an overlap ratio. Specifically, we define a local, page-utility-oriented overlap ratio at the vertex level and then aggregate it to obtain the global $OR(G)$. The overlap ratio for individual vertex $u$ quantifies the proportion of its graph neighbors that reside on the same disk page: $OR(u) = \frac{|B(u) \cap N(u)|}{n_p - 1}$\hypertarget{eq:overlap_ratio_u}{}, where $B(u)$ represents the set of vertices stored in the same disk page as vertex $u$, and $N(u)$ denotes the set of graph neighbors of vertex $u$. The denominator $n_p - 1$ excludes vertex $u$ itself, focusing on the co-location of its neighbors. For example, if a page stores $n_p{=}16$ records and $\lvert B(u)\cap N(u)\rvert{=}2$, then $OR(u)=2/(16{-}1)=13.3\%$. For a graph index $G=(V, E)$, aggregating local overlap ratios gives the global overlap ratio as the vertex-wise average: $OR(G) = \frac{\sum_{u \in V} OR(u)}{|V|}$.

\stitle{I/O complexity model.} We adopt the disk-resident best-first search model~\cite{Wang2024} to illustrate the I/O challenge. Let the disk-resident graph index be \(G=(V,E)\). Let \(H\) denote the expected hop count per query, \(\bar{R}\) the average out-degree, and \(n_p\) the number of records per page. Under a uniform page size \(P\) (\eg 4~KB) and record size \(s_{\text{rec}}\), we have \(n_p=\left\lfloor P/s_{\text{rec}}\right\rfloor\). Based on the terminology and overlap ratio defined above, the page I/O complexity for disk-resident best-first search satisfies\footnote{This model does not account for optimization techniques such as quantization, caching, and MemGraph, \etc.}
\begin{equation}
    \text{Page reads per query} = O\!\left(\frac{\bar{R} \cdot H}{OR(G) \cdot n_p}\right)
    \label{eq:io_noopt}
\end{equation}

\noindent\textbf{Intuition.} A query performs $O(\bar{R} \cdot H)$ effective neighbor expansions; each page read contributes an expected $OR(G)\cdot n_p$ useful neighbors, so the number of unique page reads scales as $O\!\big(\bar{R} \cdot H/(OR(G) \cdot n_p)\big)$. Intuitively, shorter convergence paths ($H\!\downarrow$) and higher page-level utility ($OR(G)\!\uparrow$, $n_p\!\uparrow$) reduce I/Os, while larger $\bar{R}$ expands the frontier and increases I/Os.

\subsection{Objectives}
Grounded in the theoretical model above and the profiling results in Figure~\ref{fig:io_profiling}, we systematically identify four root causes of I/O inefficiency: (1) \emph{I/O waste} occurs when graph traversal reads disk pages containing vertices that are never explored during best-first search, resulting in substantial wasted I/O operations~\cite{Wang2024}; (2) \emph{low bandwidth utilization} arises from sequential I/O--compute execution, where disk I/O remains idle during computation phases, leaving bandwidth underutilized~\cite{async-diskann,guoachieving}; (3) \emph{long search paths} emerge when entry points are far from query targets, requiring extensive graph traversal with numerous hops, each demanding independent disk access and dramatically increasing I/O operations~\cite{Subramanya2019,Wang2024,guoachieving}; and (4) \emph{frequent cache misses} occur during best-first search due to unpredictable access patterns, necessitating effective cache strategies to reduce miss rates~\cite{async-diskann,Subramanya2019,Jang2023,Jang2024}.

Through a systematic survey of existing work~\cite{Subramanya2019,Wang2024,guoachieving,async-diskann,Ni2023,Zhang2019,Zhang2018,Chen2021,BBANN,Tatsuno2025,Pan2023,Jang2023,Jang2024,Ren2020,Chen2024,Kim2023,Sun2020,Tian2024,Gao2024,Jegou2011}, we organize mainstream I/O optimization techniques into an orthogonal three-dimensional taxonomy: \textbf{Memory Layout Optimization} (\S\ref{sec:memory_layout}), \textbf{Disk Layout Optimization} (\S\ref{sec:disk_layout}), and \textbf{Search Algorithm Optimization} (\S\ref{sec:search_opt}), encompassing eight representative techniques. Table~\ref{tab:method_io_xx} maps each technique to the four bottlenecks above, revealing how different optimizations address specific I/O challenges. \S\ref{sec:design_space} provides detailed descriptions of each technique.

\stitle{Scope.} We target general-purpose SSDs and fix DiskANN's Vamana-based logical graph structure; our study focuses on physical data layout and search scheduling under realistic memory budgets. Beyond traditional SSD-based settings, specialized devices—\eg SmartSSD~\cite{Kim2023,Sun2020,Tian2024}, Persistent Memory~\cite{Ren2020,Chen2024}, and CXL~\cite{Jang2023,Jang2024,Cosmos2025}—are out of scope. While alternative logical graph structures can influence performance~\cite{Yang2024}, this work confines its scope to the physical graph structure under DiskANN's Vamana-based logic.

\section{Design Space}
\label{sec:design_space}

\begin{table}[t]
    \footnotesize 
    \renewcommand\arraystretch{0.6}
    \mycaption{0}{0}{Eight techniques versus four I/O challenges (I/O waste, bandwidth utilization, long search paths, high cache misses). \textcolor{darkgreen}{\ding{218}}: Improve; \textcolor{darkred}{\ding{216}}: cost; \textcolor{disabledgray}{\textbf{-/-}}: Reduce.}
    \label{tab:method_io_xx}
    \setlength{\tabcolsep}{0.9mm}{ 
      \begin{tabular}{|c|c|c|c|c|} 
       \hline
       \textbf{Method}                      & \textbf{I/O waste}                                                         & \textbf{\begin{tabular}[c]{@{}c@{}}Bandwidth \\ utilization\end{tabular}}  & \textbf{\begin{tabular}[c]{@{}c@{}}Long search \\ path\end{tabular}}       & \textbf{\begin{tabular}[c]{@{}c@{}}High cache \\ misses\end{tabular}}      \\ 
       \hline
       \texttt{PQ}              & \textcolor{darkgreen}{\ding{218}}             &             \textcolor{disabledgray}{\textbf{-/-}}                       &           \textcolor{disabledgray}{\textbf{-/-}}                &          \textcolor{disabledgray}{\textbf{-/-}}                   \\ 
       \hline
       \texttt{Cache}         &        \textcolor{disabledgray}{\textbf{-/-}}            &                \textcolor{disabledgray}{\textbf{-/-}}                    &            \textcolor{disabledgray}{\textbf{-/-}}               & \textcolor{darkgreen}{\ding{218}}                      \\ 
       \hline
       \texttt{MemGraph}        &           \textcolor{disabledgray}{\textbf{-/-}}         &           \textcolor{disabledgray}{\textbf{-/-}}                         & \textcolor{darkgreen}{\ding{218}}                    &           \textcolor{disabledgray}{\textbf{-/-}}                  \\ 
       \hline
       \texttt{PageShuffle}     & \textcolor{darkgreen}{\ding{218}}             &                \textcolor{disabledgray}{\textbf{-/-}}                    &            \textcolor{disabledgray}{\textbf{-/-}}               &           \textcolor{disabledgray}{\textbf{-/-}}                  \\ 
       \hline
       \texttt{All-in-Storage}  & \textcolor{darkred}{\ding{216}}             &                \textcolor{disabledgray}{\textbf{-/-}}                    &              \textcolor{disabledgray}{\textbf{-/-}}             &               \textcolor{disabledgray}{\textbf{-/-}}              \\ 
       \hline
       \texttt{DynamicWidth}    & \textcolor{darkgreen}{\ding{218}}             &            \textcolor{disabledgray}{\textbf{-/-}}                        & \textcolor{darkgreen}{\ding{218}}                    &    \textcolor{disabledgray}{\textbf{-/-}}                         \\ 
       \hline
       \texttt{Pipeline}        & \textcolor{darkred}{\ding{216}}             & \textcolor{darkgreen}{\ding{218}}                             &            \textcolor{disabledgray}{\textbf{-/-}}               &   \textcolor{disabledgray}{\textbf{-/-}}                          \\ 
       \hline
       \texttt{PageSearch}      & \textcolor{darkgreen}{\ding{218}}             &            \textcolor{darkred}{\ding{216}}                       &       \textcolor{disabledgray}{\textbf{-/-}}                    &         \textcolor{disabledgray}{\textbf{-/-}}                    \\
       \hline
       \end{tabular}
    }
   \end{table}

\begin{figure}[t]
    \centering
    \includegraphics[width=0.8\linewidth]{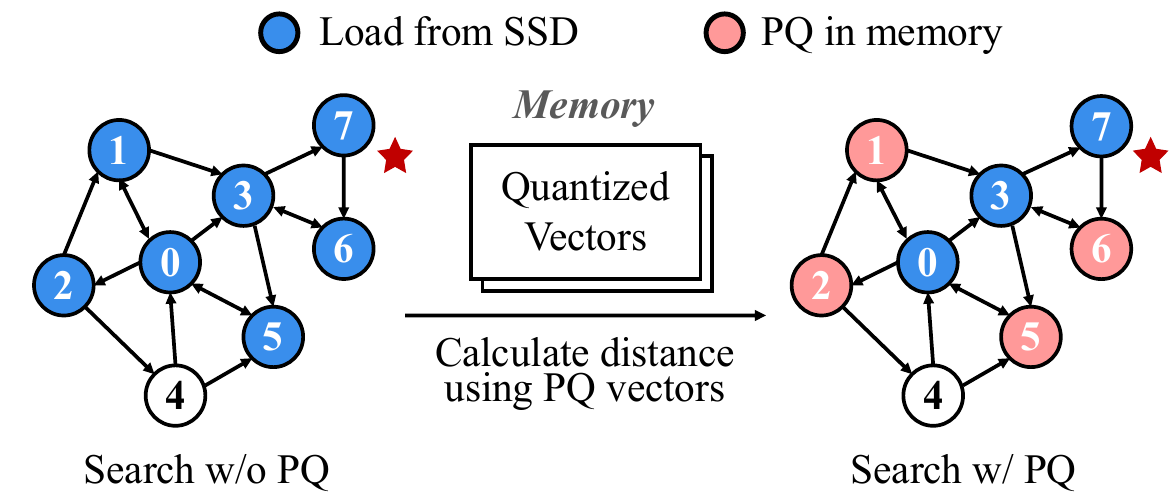}
    \mycaption{0}{-10}{Quantization reduces disk reads by moving most distance calculation to memory.}
    \label{fig:compress_vector_decrease_ioreq}
\end{figure}

\subsection{Memory Layout Optimization}\label{sec:memory_layout}
Memory layout optimization strategies reduce the frequency of disk I/O by maintaining carefully designed auxiliary data structures in memory. This dimension encompasses three approaches: (1) \emph{vector quantization} for compressing high-dimensional data~\cite{Zhang2019,Zhang2018,Subramanya2019,Jegou2011,Gao2024}; (2) \emph{cache management} for keeping frequently accessed vertices in memory~\cite{Subramanya2019,Jang2023,Jang2024}; and (3) \emph{hierarchical graphs} for reducing search path length through coarse-grained navigation~\cite{Wang2024,Ni2023,Chen2021,BBANN}.

\begin{figure*}[t]
    \centering
    \vspace{-5mm}
    \makebox[\textwidth][c]{%
        \begin{minipage}[t]{0.3\textwidth}
            \vspace{5mm}\centering
            \includegraphics[height=2.7cm]{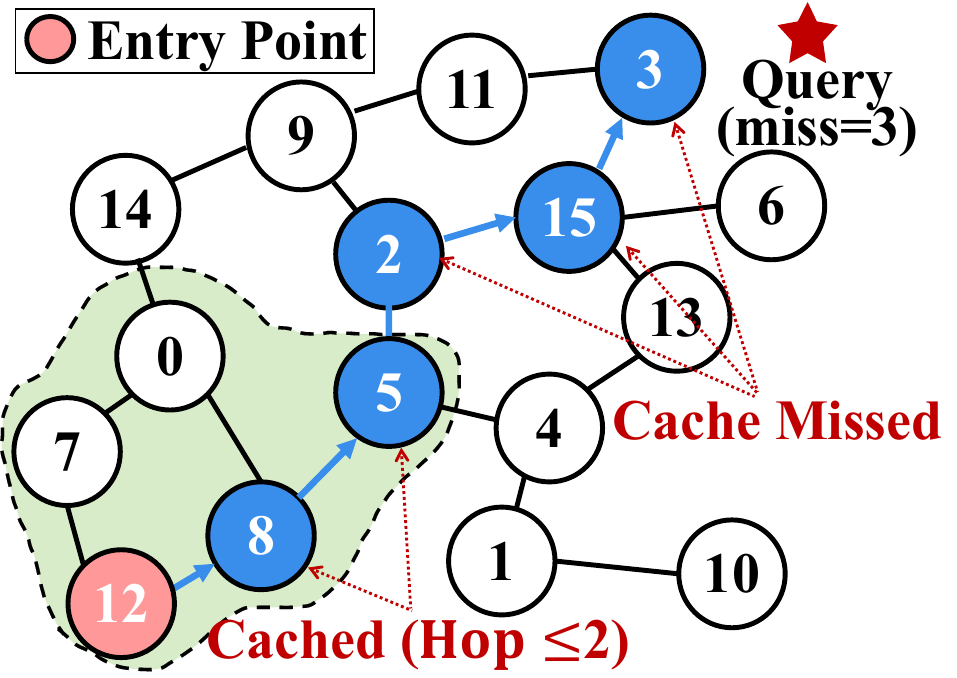}
            \setlength{\abovecaptionskip}{3mm}%
            \setlength{\belowcaptionskip}{0pt}%
            \captionof{figure}{Cache miss illustration. }
            \label{fig:cache_miss}
        \end{minipage}%
        \hfill
        \begin{minipage}[t]{0.65\textwidth}
            \vspace{0pt}\centering
            \subfloat[Entry point selection]{
                \includegraphics[height=2.7cm]{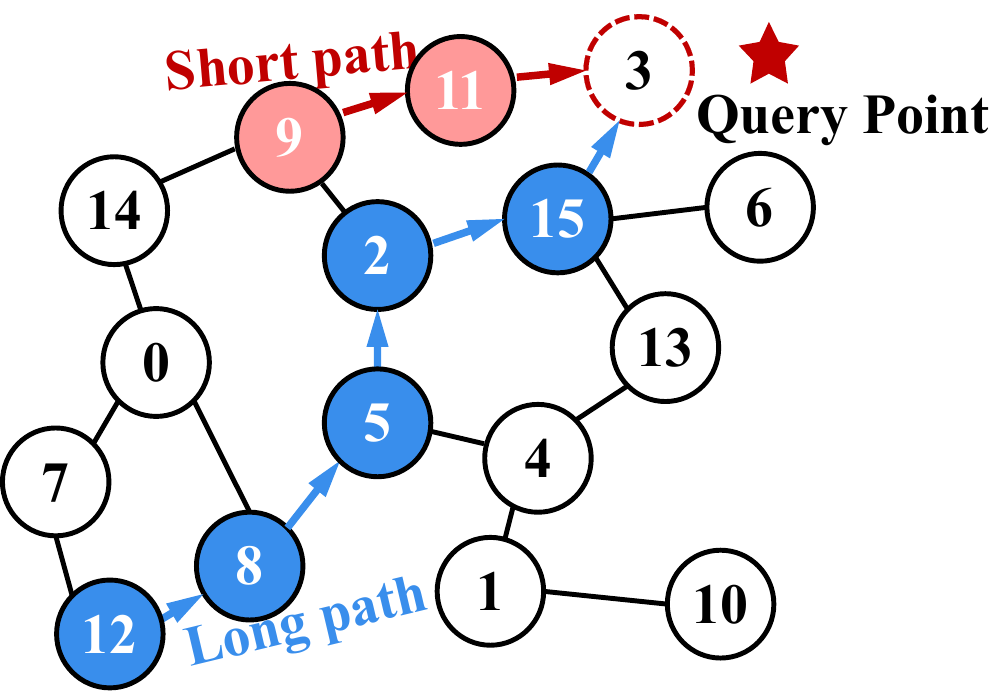}
                \label{fig:search_path}
            }
            \hfill
            \subfloat[Hierarchical graph]{
                \includegraphics[height=2.7cm]{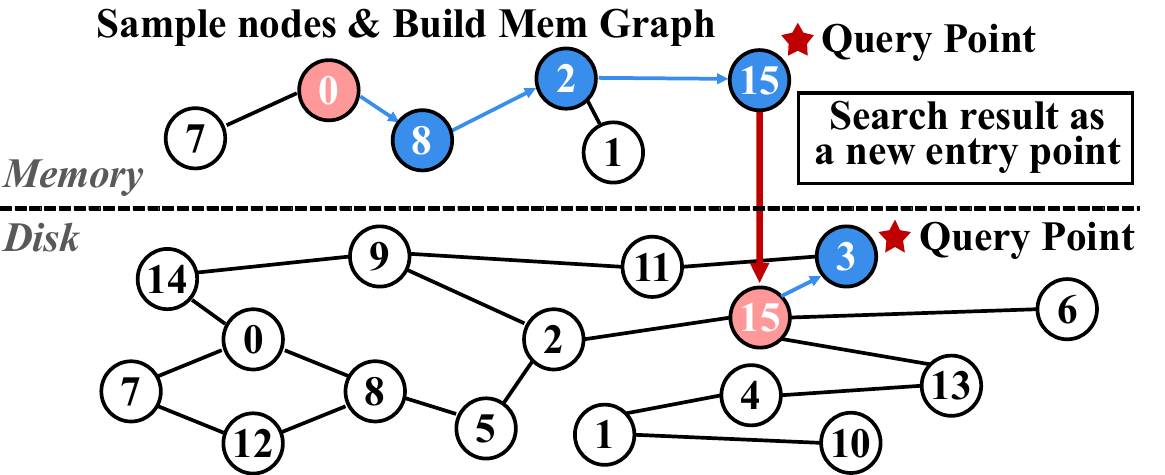}
                \label{fig:hier_graph}
            }
            \setlength{\abovecaptionskip}{0pt}%
            \setlength{\belowcaptionskip}{-10pt}%
            \captionof{figure}{Entry-point optimization. An in-memory navigation graph supplies high-quality entry points for disk-based search.}
            \label{fig:entry_point_selection}
        \end{minipage}%
    }
\end{figure*}

\subsubsection{Product Quantization (PQ)}
Product quantization~\cite{Gersho2012,Jegou2011,Ge2013} is a critical technique for achieving memory efficiency, aiming to reduce memory footprint and disk I/O by compressing high-dimensional vector representations. As shown in Figure~\ref{fig:compress_vector_decrease_ioreq}, the core benefit of quantization lies in maintaining compressed vector representations in memory (with compression rates of tens of times) while storing full-precision vectors on disk. This enables a two-stage search process: \ding{202} \ul{approximate filtering} using memory-resident quantized vectors to identify promising candidates, followed by \ding{203} \ul{precise refinement} with disk-based full vectors only for top candidates. The key insight is that distance calculations can be performed directly using quantized coordinates in memory, eliminating the need to read full vectors from disk for initial candidate ranking. In the illustrated example (Figure~\ref{fig:compress_vector_decrease_ioreq}), traditional DiskANN requires 7 disk I/O operations for vector distance computations, while the quantization-optimized approach reduces this to only 3 I/O operations primarily for neighbor topology retrieval, achieving significant I/O efficiency improvement. 
Formally, with PQ optimization, the I/O complexity from Equation~\ref{eq:io_noopt} is reduced to:
\begin{equation}
    \text{Optimized page reads} = O\left(\frac{H}{OR(G) \cdot n_p}\right)
    \label{eq:io_pq}
\end{equation}
This improvement eliminates the $\bar{R}$ factor in the numerator because distance computations are performed in memory and no longer require disk reads. Theoretically, the lower bound for page reads is $O(H / n_p)$ (i.e., all required vectors are co-located on the same page). However, this ideal page locality is difficult to attain in practice due to the inherent randomness of graph traversals and the complexity of optimizing physical layouts for dynamic query patterns.

However, vector quantization introduces precision loss, creating a trade-off between memory efficiency and search accuracy. To mitigate this limitation, typical approaches expand the candidate set during search and perform precise re-ranking using full vectors from disk for final Top-k results. DiskANN's co-location design stores vector data and neighbor information together, ensuring that reading neighbor IDs for candidate expansion simultaneously retrieves associated vector data, avoiding additional disk I/Os for the re-ranking phase.

Vector quantization methods vary significantly in accuracy; we focus on classic Product Quantization (PQ)~\cite{Jegou2011} as the representative quantization technique. We experimentally evaluate quantization's impact on I/O efficiency and search performance in DiskANN systems.

\subsubsection{Cache Management (Cache)}
Efficient cache management is crucial for reducing disk I/O and improving DiskANN's query performance. Similar to hierarchical index structures like B+ trees~\cite{DBS-028,202263}, where near root nodes are accessed more frequently, nodes closer to search entry points in graph traversal exhibit higher access probability. DiskANN's current caching strategy employs Single-Source Shortest Path (SSSP)~\cite{Dijkstra1959} algorithm to pre-load nodes within certain hops from the search entry point (Figure~\ref{fig:cache_miss}). However, this static approach may not adapt well to diverse query patterns and varying access distributions across datasets.

Various caching approaches exist with different characteristics: (1) \emph{SSSP caching}: Pre-loads vertices within a fixed number of hops from entry points; (2) \emph{Frequency-based caching}~\cite{Wang2024}: Dynamically tracks and caches vertices with the highest access frequencies; (3) \emph{LRU caching}~\cite{Tsao1972}: Maintains recently accessed vertices based on temporal locality assumptions. We focus on the widely adopted SSSP caching strategy in our experimental evaluation.

\subsubsection{Hierarchical Graphs (MemGraph)}
As established in Figure~\ref{fig:search_path}, entry point quality fundamentally determines search path length and I/O efficiency. While DiskANN employs static entry point selection, this naive approach fails to leverage the geometric structure of high-dimensional vector spaces. Hierarchical graph structures address this limitation by implementing a two-level search strategy: a fast in-memory coarse-grained navigation followed by precise disk-based refinement~\cite{Wang2024,Ni2023,guoachieving,Zhang2019,Zhang2018,Chen2021,BBANN}. The hierarchical approach, illustrated in Figure~\ref{fig:hier_graph}, maintains a memory-resident subgraph constructed from strategically sampled vertices. During query processing, the algorithm first traverses this in-memory structure to identify promising regions, then uses the discovered vertices as high-quality entry points for disk-based search. This design exploits the principle that effective entry points should be geometrically close to query targets, significantly reducing the expected number of disk I/O operations required for convergence.

The effectiveness of hierarchical graphs critically depends on two key design choices: sampling ratio and sampling strategy. The sampling ratio determines the trade-off between memory overhead and search path reduction—insufficient sampling fails to provide adequate coverage for effective entry point selection, while excessive sampling introduces substantial memory overhead and prolongs in-memory search phases. Different sampling strategies exist, including \emph{random sampling}~\cite{Wang2024}, which provides uniform coverage, and \emph{clustering-based sampling}~\cite{Ni2023}, which selects cluster centroids for better geometric representation. We employ standard random sampling in our experimental implementation.

\subsection{Disk Layout Optimization}\label{sec:disk_layout}

This section discusses optimization methods for different physical storage structures under the same logical graph topology. Varying storage structures can significantly impact the I/O access process. We will primarily discuss the following two storage structures:

\begin{figure}[t]
    \centering
    \subfloat[Origin Page Layout]{
        \includegraphics[width=0.48\linewidth]{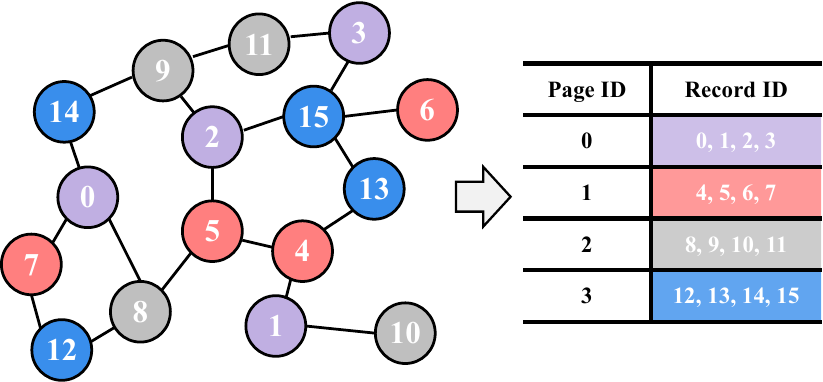}
        \Description{Origin page layout.}
        \label{fig:origin_page_layout}
    }
    \hfill
    \subfloat[Page Shuffle]{
        \includegraphics[width=0.48\linewidth]{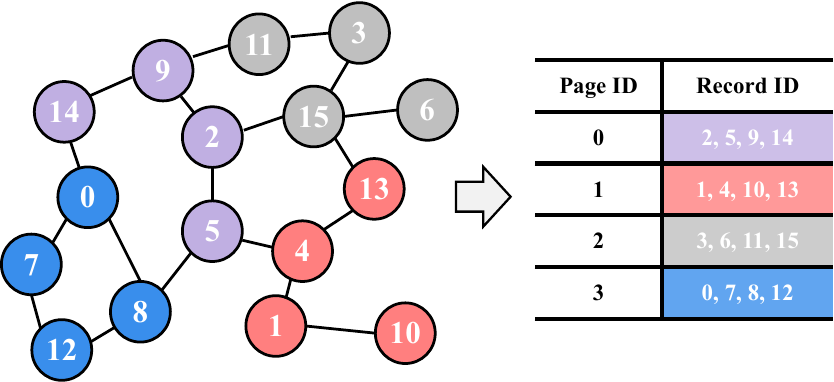}
        \Description{Page shuffle.}
        \label{fig:page_shuffle}
    }
    \mycaption{0}{0}{Page shuffle groups neighboring vertices onto the same page to improve locality.}
    \label{fig:disk_layout_optimization}
\end{figure}

\subsubsection{Page Shuffle (PS)}

The fundamental I/O efficiency challenge identified in \S\ref{sec:challenges} stems from poor data locality~\cite{liang2019cognitive}—traditional DiskANN's vertex-ID-based sequential storage scatters graph neighbors across different disk pages (\cf Figure~\ref{fig:origin_page_layout}), resulting in extremely low overlap ratios (\eg 6.25\% on SIFT1B dataset~\cite{Wang2024}). Page shuffle addresses this core limitation by reorganizing physical data placement to maximize overlap ratio during graph traversal.
Starling~\cite{Wang2024} introduces page shuffle to transform scattered neighbor distributions into locality-aware arrangements. As illustrated in Figure~\ref{fig:page_shuffle}, the technique ensures that each disk I/O operation serves multiple navigation steps by grouping graph neighbors within the same pages. This dramatically improves the vertex overlapping ratio (Eq.~\ref{eq:io_noopt}).

\subsubsection{All-in-Storage Layout (AiS)}

\begin{figure}[t]
    \centering
    \includegraphics[width=0.7\linewidth]{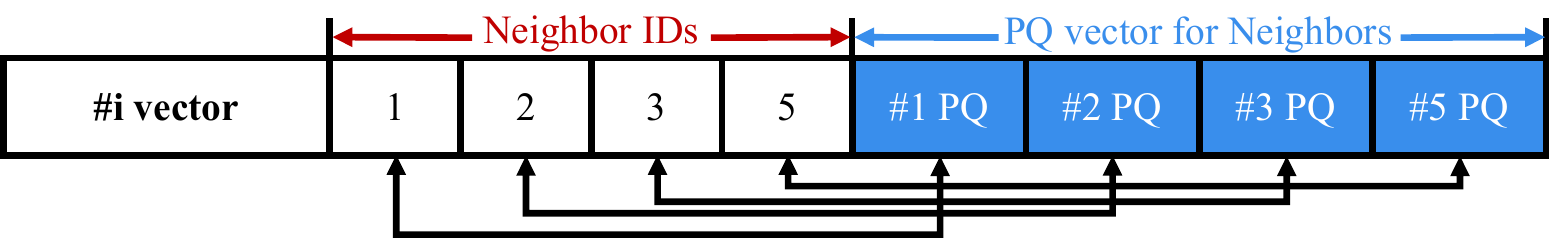}
    \Description{AiSAQ.}
    \mycaption{0}{0}{AiSAQ layout. Each record stores the node vector and neighbor IDs, plus PQ codes of neighbors.}
    \label{fig:AiSAQ}
\end{figure}

Alternatively, an all-in-storage layout stores both full-precision vectors and PQ codes on SSD, thereby reducing the memory footprint. All-in-Storage ANN with Product Quantization (AiSAQ)~\cite{Tatsuno2025} exemplifies this design by moving full-precision vectors and PQ codes from memory to storage. As shown in Figure~\ref{fig:AiSAQ}, AiSAQ co-locates vector data, neighbor IDs, and the PQ codes of neighboring vertices within the same storage units.
This design reduces memory requirements from gigabytes to merely megabytes for billion-scale datasets while maintaining competitive search performance. The approach offers significant advantages for memory-constrained environments and enables rapid switching between multiple large-scale indexes—a crucial capability for applications like retrieval-augmented generation (RAG) that require dynamic switching between multiple indices~\cite{RAGCache}. 
However, we do not include AiSAQ in our experimental evaluation, as it targets memory-constrained scenarios with fundamentally different design trade-offs. Comparing AiSAQ against methods that utilize more substantial memory budgets (as in our evaluation setup) would be unfair, since AiSAQ's primary optimization goal—minimizing memory footprint—conflicts with our focus on I/O efficiency under realistic memory budgets.

\subsection{Search Algorithm Optimization}\label{sec:search_opt}

Beyond optimizing data layout, search algorithm optimization focuses on improving the execution efficiency of graph traversal itself. This dimension addresses two critical algorithmic bottlenecks: (1) \emph{vertex utilization inefficiency} when accessing disk pages, and (2) \emph{bandwidth underutilization} caused by sequential I/O--compute execution. These strategies complement layout improvements by maximizing the effectiveness of each disk access operation.

\begin{figure}[t]
    \centering
    \includegraphics[width=0.6\linewidth]{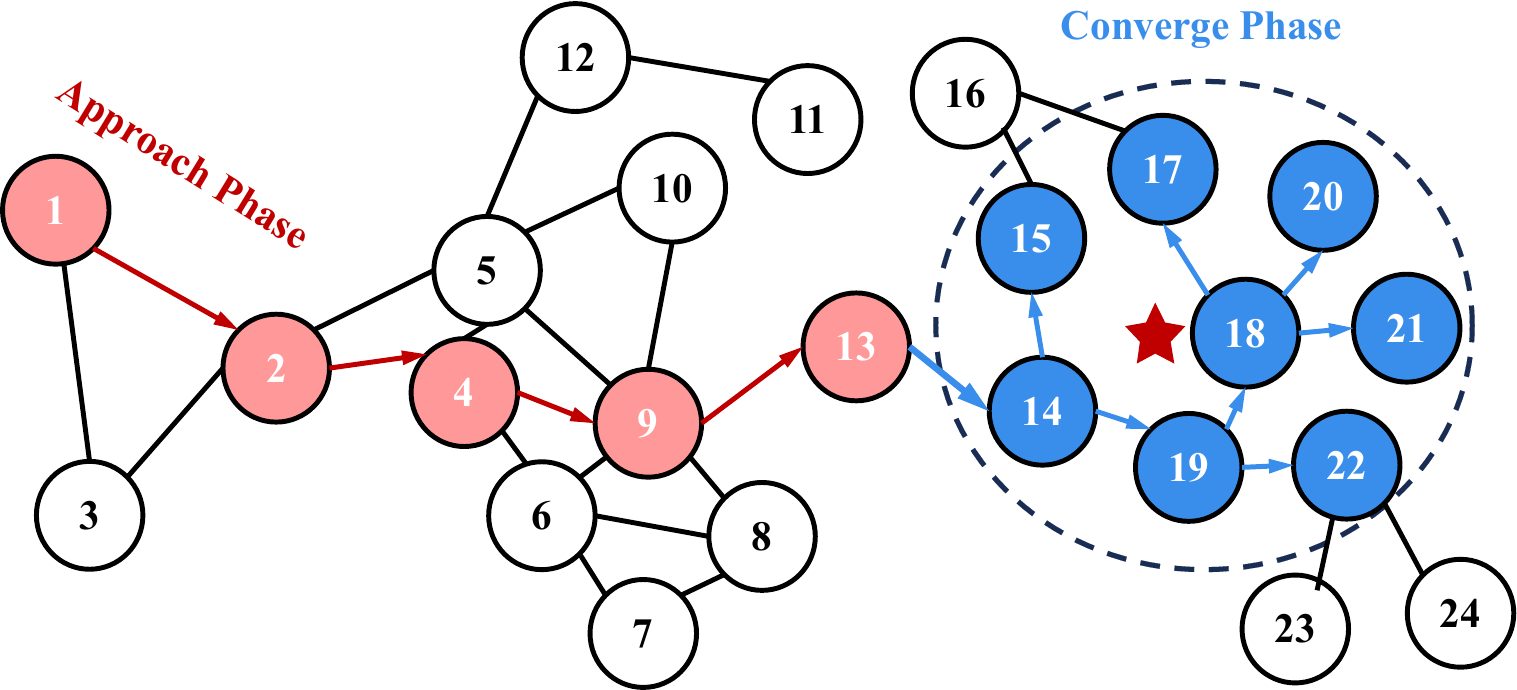}
    \Description{Dynamic search width.}
    \mycaption{0}{-10}{Dynamic search width. The beam width increases progressively across the approach and converge phases.}
    \label{fig:dynamic_search_width}
\end{figure}

\subsubsection{Dynamic Width (DW)}
The best-first search or beam search process can be naturally divided into two distinct phases: the \emph{approach phase} and the \emph{converge phase}. As illustrated in Figure~\ref{fig:dynamic_search_width}, during the \ding{202} \ul{approach phase}, the search rapidly navigates toward the vicinity of the target query vector. In the \ding{203} \ul{converge phase}, the search gradually converges to the target query vector. This two-phase phenomenon has been discussed extensively in multiple studies~\cite{guoachieving,288618,Yin2025}. PipeANN~\cite{guoachieving} leverages the progressive convergence characteristics of beam search by adopting a smaller search width during the \textit{approach phase} to avoid unnecessary I/O operations, since larger search widths do not improve recall effectiveness in this phase. Subsequently, the search width is gradually increased during the \emph{converge phase} to accelerate the convergence process and improve search efficiency.
To make this intuition precise, we define the \emph{I/O utilization} as
\begin{equation}
U_{\mathrm{io}} \triangleq \frac{N_{\mathrm{eff}}}{N_{\mathrm{read}}}
= \frac{N_{\mathrm{eff}}}{N_{\mathrm{eff}} + N_{\mathrm{rbu}}},
\label{eq:io_utilization}
\end{equation}
where \(N_{\mathrm{read}}\) is the number of nodes retrieved from SSD, \(N_{\mathrm{eff}}\) is the number of retrieved nodes that are actually explored (expanded) by the search, and \(N_{\mathrm{rbu}}\) denotes ``read-but-unexplored'' nodes. Increasing beam width \(\omega\) tends to inflate \(N_{\mathrm{rbu}}\) (especially in the approach phase where extra reads rarely improve recall), thus lowering \(U_{\mathrm{io}}\) and wasting I/O. Dynamic width mitigates this by keeping \(\omega\) small early to suppress wasted reads, and gradually increasing \(W\) later when more retrieved candidates are likely to be explored, improving both I/O utilization and convergence.

\begin{table*}[t]
    \small 
    \renewcommand\arraystretch{0.7}
    \mycaption{0}{0}{Coverage of representative systems across the three optimization dimensions.}
       \label{tab:algorithm_optimization_comparison}
    \setlength{\tabcolsep}{1.6mm}{ 
       \begin{tabular}{|l|c|c|c|c|c|c|c|c|}
       \hline
       \multirow{2}{*}{\textbf{Algorithms}} & \multicolumn{3}{c|}{\textbf{Memory Layout Optimization}} & \multicolumn{2}{c|}{\textbf{Disk Layout Optimization}} & \multicolumn{3}{c|}{\textbf{Search Algorithm Optimization}} \\
       \cline{2-9}
       & \texttt{PQ} & \texttt{Cache} & \texttt{MemGraph} & \texttt{PageShuffle} & \texttt{All-in-Storage} & \texttt{DynamicWidth} & \texttt{Pipeline} & \texttt{PageSearch} \\
       \hline
       DiskANN~\cite{Subramanya2019} & \textcolor{darkgreen}{\ding{51}} & \textcolor{darkgreen}{\ding{51}} & \textcolor{darkred}{\ding{55}} & \textcolor{darkred}{\ding{55}}& \textcolor{darkred}{\ding{55}}&\textcolor{darkred}{\ding{55}} & \textcolor{darkred}{\ding{55}} & \textcolor{darkred}{\ding{55}} \\
       \hline
       Starling~\cite{Wang2024}/DiskANN++~\cite{Ni2023} &\textcolor{darkgreen}{\ding{51}} & \textcolor{darkred}{\ding{55}} & \textcolor{darkgreen}{\ding{51}} & \textcolor{darkgreen}{\ding{51}} & \textcolor{darkred}{\ding{55}} & \textcolor{darkred}{\ding{55}} & \textcolor{darkred}{\ding{55}} & \textcolor{darkgreen}{\ding{51}} \\
       \hline
       AiSAQ~\cite{Tatsuno2025}/LM-DiskANN~\cite{Pan2023} & \textcolor{darkgreen}{\ding{51}}& \textcolor{darkred}{\ding{55}} & \textcolor{darkred}{\ding{55}} & \textcolor{darkred}{\ding{55}} & \textcolor{darkgreen}{\ding{51}} & \textcolor{darkred}{\ding{55}} & \textcolor{darkred}{\ding{55}} & \textcolor{darkred}{\ding{55}} \\
       \hline
       PipeANN~\cite{guoachieving} &\textcolor{darkgreen}{\ding{51}} & \textcolor{darkred}{\ding{55}} & \textcolor{darkgreen}{\ding{51}} & \textcolor{darkred}{\ding{55}} & \textcolor{darkred}{\ding{55}} & \textcolor{darkgreen}{\ding{51}} & \textcolor{darkgreen}{\ding{51}} & \textcolor{darkred}{\ding{55}} \\
       \hline
       GRIP~\cite{Zhang2019}/Zoom~\cite{Zhang2018} & \textcolor{darkgreen}{\ding{51}} & \textcolor{darkred}{\ding{55}} & \textcolor{darkgreen}{\ding{51}} & \textcolor{darkred}{\ding{55}} & \textcolor{darkred}{\ding{55}} & \textcolor{darkred}{\ding{55}} & \textcolor{darkred}{\ding{55}} & \textcolor{darkred}{\ding{55}} \\
       \hline
       SPANN~\cite{Chen2021}/BBANN~\cite{BBANN} & \textcolor{darkred}{\ding{55}} & \textcolor{darkred}{\ding{55}} & \textcolor{darkgreen}{\ding{51}} & \textcolor{darkgreen}{\ding{51}} & \textcolor{darkred}{\ding{55}} & \textcolor{darkred}{\ding{55}} & \textcolor{darkred}{\ding{55}} & \textcolor{darkred}{\ding{55}} \\
       \hline
       \end{tabular}
    }
\end{table*}

\subsubsection{Pipeline Search (Pipeline)}
\begin{figure}[t]
    \centering
    \vspace{-8pt}
    \subfloat[Overlap I/O and Compute]{    
        \includegraphics[width=0.52\linewidth]{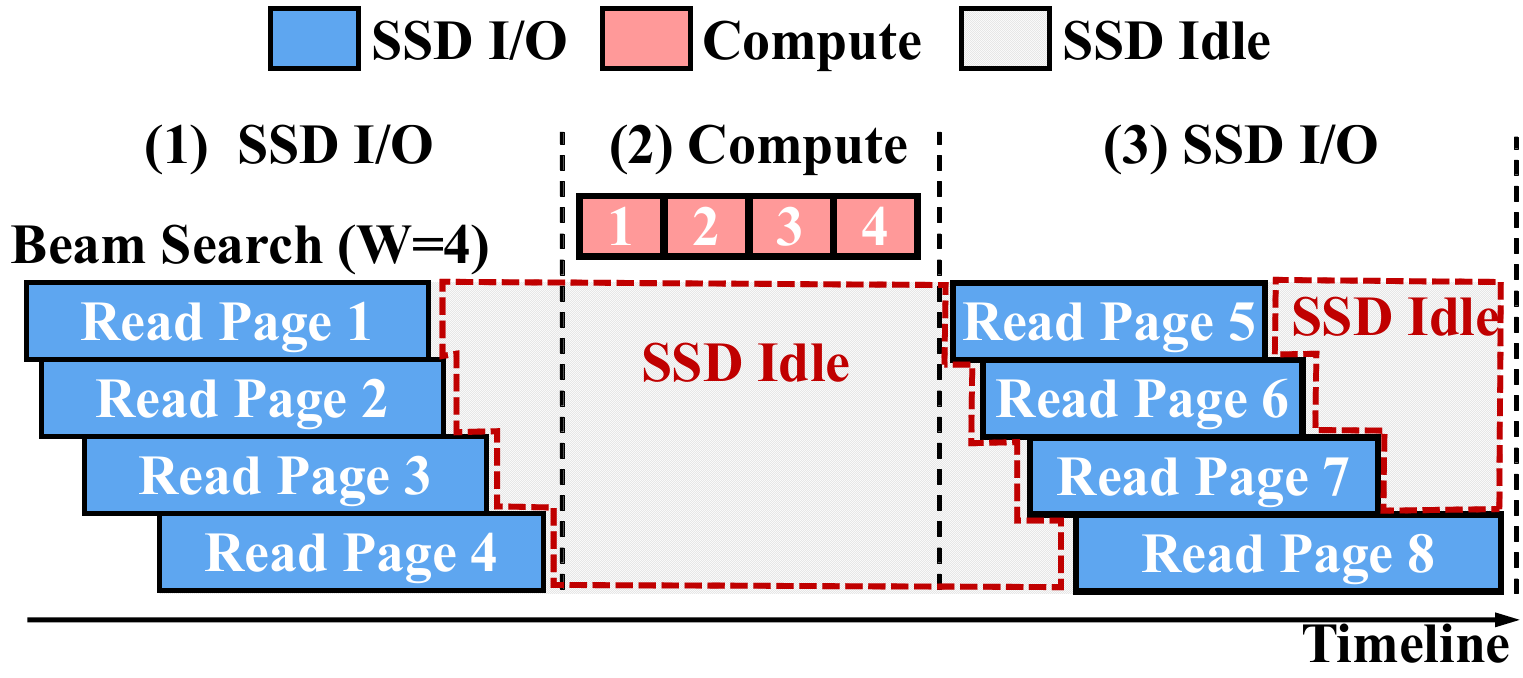}
        \label{fig:overlap_io_and_compute}
    }
    \subfloat[Pipeline I/O]{
        \raisebox{7pt}{\includegraphics[width=0.46\linewidth]{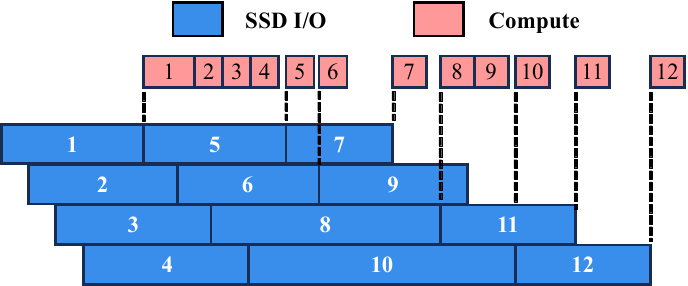}}
        \label{fig:pipeline_io}
    }
    \Description{Pipeline Processing.}
    \mycaption{0}{-15}{Pipelined processing: (a) overlap compute and I/O to improve utilization; (b) continuous I/O pipelining to maximize bandwidth efficiency.}
    \label{fig:pipeline_processing}
\end{figure}

Traditional disk-based graph search suffers from severe bandwidth underutilization due to sequential execution patterns. As illustrated in Figure~\ref{fig:overlap_io_and_compute}, beam search's alternating I/O and computation phases leave storage devices idle during processing, resulting in actual bandwidth utilization far below hardware capabilities.

Pipeline search addresses this limitation through two progressive optimizations. First, asynchronous I/O-compute overlap~\cite{async-diskann} enables simultaneous execution of distance calculations and disk reads, improving resource utilization. Second, continuous I/O pipelining~\cite{guoachieving} eliminates batch processing delays by immediately issuing new I/O requests upon completion of individual operations, as demonstrated in Figure~\ref{fig:pipeline_io}. This approach maintains near-continuous disk activity, achieving high bandwidth utilization and significantly reducing query latency.

\subsubsection{Page Search (PSe)}
While pipeline search optimizes bandwidth utilization, page search addresses vertex utilization inefficiency within loaded disk pages. Given that distance computation overhead is orders of magnitude lower than disk I/O latency, page search exploits this asymmetry by computing distances to all vertices within each accessed page, not just the target neighbors.

This approach transforms the fundamental trade-off in disk-based search: instead of minimizing distance calculations, page search maximizes the utility of expensive disk accesses. When retrieving a neighbor during graph traversal, the algorithm evaluates all colocated vertices within the same page, potentially discovering closer candidates without additional I/O operations. This strategy effectively increases the number of search nodes without proportional I/O overhead, improving search quality while maintaining I/O efficiency. The key insight is that leveraging the full content of each disk page can enhance search effectiveness at minimal computational cost.

\subsection{Algorithmic Landscape Analysis}
\label{sec:algorithmic_landscape}

Grounded in our three-dimensional taxonomy, we position representative systems against the eight techniques summarized in Table~\ref{tab:algorithm_optimization_comparison}. This mapping clarifies each system's design center and expected I/O behaviors within the design space. In short, recent progress has moved from single-axis tweaks to multi-dimensional, synergy-driven designs that directly address the four I/O bottlenecks identified earlier.

\section{Experimental Testbed}
\label{sec:experiments}

This section describes our experimental testbed and evaluation methodology, including hardware/software settings, datasets, metrics, and implementation details. 

\subsection{Experimental Settings}
\textbf{Hardware.} 
We run experiments on a server with an Intel Xeon w7-3455 (24 cores), 128~GB DDR4 memory, and a 4~TB NVMe SSD, on Ubuntu~22.04. Using \texttt{fio}~\cite{fio}, we measure 4KB random read IOPS of \textbf{819 K} and random read bandwidth of \textbf{3{,}200 MB/s}, representative of high-performance SSDs. We also measure 16KB random read IOPS of \textbf{318 K} and random read bandwidth of \textbf{4,962 MB/s}.

\noindent\textbf{Datasets.} We evaluate on four widely adopted public datasets, SIFT~\cite{SIFT}, DEEP~\cite{babenko2016efficient}, SPACEV~\cite{SPACEV1B}, and GIST~\cite{SIFT}, covering natural clustering, deep embeddings, and production-scale data. Dataset statistics and index build parameters are summarized in Table~\ref{tab:datasets}.

\noindent\textbf{Performance indicators.} We report search efficiency (queries per second, QPS; mean per-query latency), accuracy (Recall@10), space footprint (disk and memory), and I/O metrics (I/O operations per query, I/O operations per second, IOPS; bandwidth in MB/s). Unless otherwise stated, comparisons are made at matched Recall@10. 
We do not separately report Eq.~\ref{eq:io_utilization}, as it is strongly correlated with the number of I/O operations per query.

\begin{table}[t]
    \small 
    \renewcommand\arraystretch{0.8}
    \mycaption{0}{0}{Datasets and parameters used in the evaluation. $B$ and $M$ are memory budgets (GB).}
       \label{tab:datasets}
    \setlength{\tabcolsep}{0.9mm}{ 
        \begin{tabular}{|c|c|c|c|c|c|c|c|c|c|} 
       \cline{2-10}
       \multicolumn{1}{l|}{} & \multicolumn{4}{c|}{\textbf{Dataset Statistics}}                & \multicolumn{5}{c|}{\textbf{Index-Build Param.}}  \\
       \hline
       \multicolumn{1}{|c|}{\textbf{Dataset}}         & \textbf{Dim} & \textbf{Type} & \textbf{\#Vec} & \textbf{\#Q} & \textbf{R}  & \textbf{L}   & $\bm{\alpha}$ & \textbf{B}    & \textbf{M}  \\
       \hline
       \textbf{SIFT}~\cite{SIFT}             & 128          & uint8         & 100M        & 10K             & 64 & 125 & 1.2 & 4.0    & 50                            \\ 
       \cline{1-1}
       \textbf{DEEP}~\cite{babenko2016efficient}             & 96           & float         & 100M        & 10K             & 64 & 125 & 1.2 & 4.0    & 50                            \\ 
       \cline{1-1}
       \textbf{SPACEV}~\cite{SPACEV1B}           & 100          & int8         & 100M        & 29K             & 64 & 125 & 1.2 & 3.0    & 40                            \\
       \cline{1-1}
       \textbf{GIST}~\cite{SIFT}               & 960          & float         & 1M          & 10K             & 64 & 125 & 1.2 & 0.2 & 3                            \\ 
       \hline
       \end{tabular}
    }
    \vspace{-8pt}
   \end{table}

\noindent\textbf{Implementation and parameters.} We build on DiskANN~\cite{diskann-github} and integrate representative optimizations from Starling~\cite{starling-github} and PipeANN~\cite{pipeann-github}. As io\_uring is a system-level optimization that all algorithmes can be applied but currently only PipeANN use it.  To ensure fairness, we disable io\_uring~\cite{iouring} in PipeANN so that all methods use the same asynchronous I/O engine (\texttt{libaio}). 
Note that all the I/O operations are performed in DIRECT\_IO mode, \ie without using the file system page cache.
We run with 48 workers; device utilization (IOPS and bandwidth) is sampled at 1~s intervals via \texttt{iostat}. For index construction, we adopt product quantization under a memory budget $B$, partition data via overlapping clustering, and build per-shard indexes under budget $M$ using Vamana parameters: maximum out-degree $R=64$, candidate set size $L=125$, and pruning with $\alpha=1.2$; shards are then merged and stored with a 4~KB page size (we use 8~KB or 16~KB pages for GIST due to its larger per-record size). During query processing, we use a beam width $\omega=8$ with dynamic-width expansion up to $\omega=32$. For MemGraph, we sample 0.1\% vertices to construct an in-memory navigation graph ($R=48$, $L=128$) and employ a Single-Source Shortest Path (SSSP)-based cache with 0.1\% records. Results are averaged over five independent runs. Dataset-specific budgets and settings are summarized in Table~\ref{tab:datasets}.

\begin{figure}[t]
    \centering
    \vspace{-2mm}
    \includegraphics[width=\linewidth]{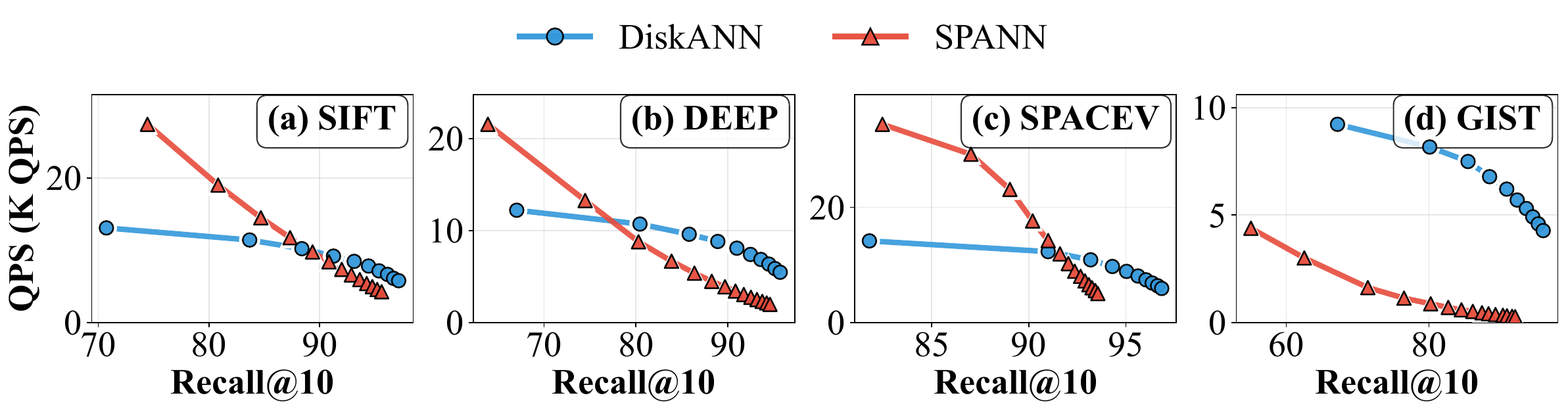}
    \Description{Performance comparison of SPANN and DiskANN on four datasets.}
    \mycaption{0}{-15}{Performance comparison of SPANN and DiskANN on four datasets.}
    \label{fig:spann_vs_diskann_comparison}
\end{figure}

\begin{figure*}[t]
    \centering
    \includegraphics[width=0.98\linewidth]{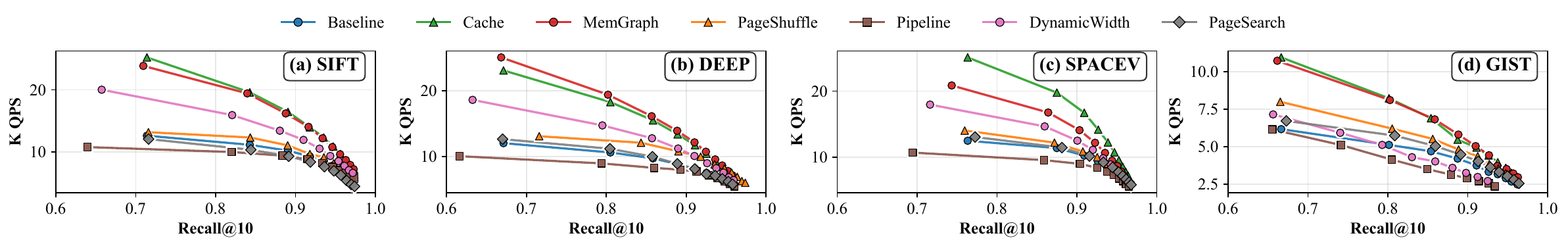}
    \mycaption{0}{-9}{Recall--QPS trade-off: seven optimizations across four datasets.}
    \label{fig:recall_qps_comparison}
\end{figure*}

\noindent\textbf{Scope of evaluation.} We do not include billion-scale datasets. The reason is that \texttt{PageShuffle} requires sizeable adjacency information and the page-reordering structure to reside in memory during offline shuffling; its peak memory usage grows near-linearly with data size. On our 128\,GB server, the largest graph we can reliably build is about 100M scale (\cf Table~\ref{tab:constrction_overhead_comparison}). Empirically, at the 100M data scale, our runs saturate SSD random-read IOPS and bandwidth (\cf Table~\ref{tab:disk_performance_comparison} and Table~\ref{tab:combination_resource_utilization}), placing the system squarely in the I/O-bound regime and enabling stable differentiation among methods. Importantly, the 100M scale fully exercises the dominant I/O bottlenecks of disk-based ANN and reliably differentiates methods in terms of throughput, latency, and read amplification, thereby sufficing to support the efficiency comparisons and conclusions of this paper. A systematic billion-scale evaluation is deferred to future work on higher-memory platforms together with engineering improvements (\eg out-of-memory page reordering).

\subsection{Graph-based vs. Inverted Indices}
Although this paper primarily focuses on evaluating graph-based indexing techniques, we first investigate whether graph indices consistently outperform inverted indices. We evaluated SPANN~\cite{Chen2021} following SPFresh~\cite{xu2023spfresh} parameter settings; note that we enable the replica feature and set the replication factor to 8. As shown in Figure~\ref{fig:spann_vs_diskann_comparison}, for low-dimensional data, SPANN's performance degrades much more rapidly than DiskANN's as the recall target increases. This is because achieving higher recall requires SPANN to read coarse-grained posting lists, leading to more I/O operations and consequently faster performance degradation. Furthermore, on high-dimensional datasets, SPANN exhibits poor performance due to the curse of dimensionality, which significantly increases the difficulty of effective clustering. Consequently, SPANN outperforms DiskANN only on low-dimensional datasets under low recall requirements. Therefore, we decided to exclude SPANN from our subsequent detailed evaluation.

Beyond the search performance, we also compare the overhead of index construction for both index methods. As shown in Table~\ref{tab:spann_diskann_construction_comparison}, index construction for both SPANN and DiskANN is very time-consuming, but SPANN's index size is 1.5× to 3.4× larger than DiskANN's. Because SPANN's posting lists contain substantial redundant vector information across different lists, resulting in significant storage space amplification. During search, SPANN exhibits higher SSD bandwidth but lower IOPS, as its basic I/O granularity is coarser, \ie posting lists are typically larger than page size.
\begin{center}
    \fbox{
    \begin{minipage}{0.95\linewidth}
\textbf{Finding 1:} \emph{SPANN outperforms DiskANN only on low-dimensional datasets under low recall requirements. On high-dimensional datasets, SPANN exhibits poor performance due to the curse of dimensionality, which significantly increases the difficulty of effective clustering. Moreover, SPANN's index size is 1.5× to 3.4× larger than DiskANN's.}
    \end{minipage}
    }
\end{center}

\begin{table}[t]
    \footnotesize
    \mycaption{2}{-10}{Index construction overhead and disk I/O performance comparison between SPANN and DiskANN.}
    \label{tab:spann_diskann_construction_comparison}
    \renewcommand\arraystretch{0.8}
    \setlength{\tabcolsep}{0.4mm}{ 
        \begin{tabular}{|l|l|cc|cc|}
            \hline
                                               &                                      & \multicolumn{2}{c|}{\textbf{Index construction Overhead}}                                      & \multicolumn{2}{c|}{\textbf{Search I/O Performance}}                           \\ \cline{3-6} 
            \multirow{-2}{*}{\textbf{Dataset}} & \multirow{-2}{*}{\textbf{Algorithm}} & \multicolumn{1}{c|}{\textbf{Index Size(GB)}}      & \multicolumn{1}{l|}{\textbf{Build Time(s)}} & \multicolumn{1}{c|}{\textbf{IOPS}}               & \textbf{Bandwidth (MB/s)} \\ \hline
                                               & \textbf{SPANN}                       & \multicolumn{1}{c|}{\cellcolor[HTML]{F6C1BD}84.4}  & 8803                                     & \multicolumn{1}{c|}{478}                         & 4843                      \\ \cline{2-6} 
            \multirow{-2}{*}{\textbf{SIFT}}    & \textbf{DiskANN}                     & \multicolumn{1}{c|}{39.1}                          & 8814                                     & \multicolumn{1}{c|}{\cellcolor[HTML]{F6C1BD}791} & 3091                      \\ \hline
                                               & \textbf{SPANN}                       & \multicolumn{1}{c|}{\cellcolor[HTML]{F6C1BD}232.1} & 12033                                    & \multicolumn{1}{c|}{223}                         & 5425                      \\ \cline{2-6} 
            \multirow{-2}{*}{\textbf{DEEP}}    & \textbf{DiskANN}                     & \multicolumn{1}{c|}{67.3}                          & 14960                                    & \multicolumn{1}{c|}{\cellcolor[HTML]{F6C1BD}819} & 3200                      \\ \hline
                                               & \textbf{SPANN}                       & \multicolumn{1}{c|}{\cellcolor[HTML]{F6C1BD}60.1}  & 8988                                     & \multicolumn{1}{c|}{561}                         & 4570                      \\ \cline{2-6} 
            \multirow{-2}{*}{\textbf{SPACEV}}  & \textbf{DiskANN}                     & \multicolumn{1}{c|}{39.3}                          & 6139                                     & \multicolumn{1}{c|}{\cellcolor[HTML]{F6C1BD}768} & 3001                      \\ \hline
                                               & \textbf{SPANN}                       & \multicolumn{1}{c|}{\cellcolor[HTML]{F6C1BD}12.4}  & 520                                      & \multicolumn{1}{c|}{303}                         & 5064                      \\ \cline{2-6} 
            \multirow{-2}{*}{\textbf{GIST}}    & \textbf{DiskANN}                     & \multicolumn{1}{c|}{7.7}                           & \cellcolor[HTML]{F6C1BD}1174             & \multicolumn{1}{c|}{\cellcolor[HTML]{F6C1BD}512}                         & 3997                      \\ \hline
            \end{tabular}
    }
    \vspace{-5mm}
\end{table}

\begin{figure*}[t]
    \centering
    \includegraphics[width=\linewidth]{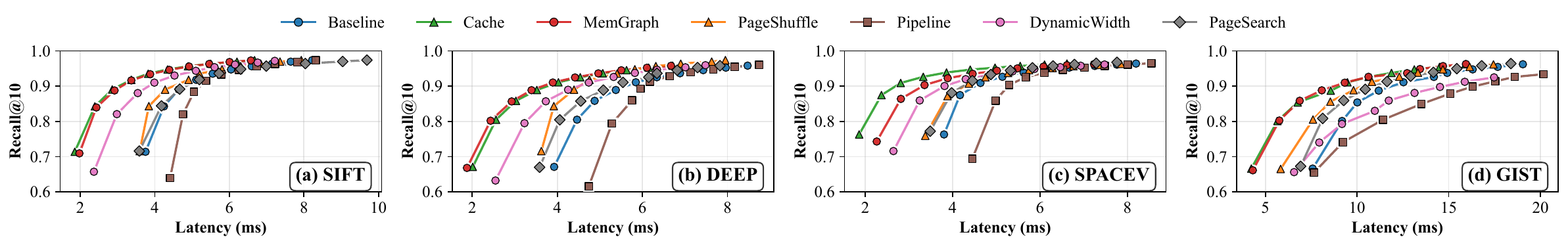}
    \Description{Latency versus Recall comparison across seven optimization algorithms and four datasets, showing performance-accuracy relationships.}
    \mycaption{0}{-15}{Latency--Recall@10 trade-off across seven optimizations and four datasets.}
    \label{fig:latency_recall_comparison}
\end{figure*}

\begin{figure*}[t]
    \centering
    \includegraphics[width=\linewidth]{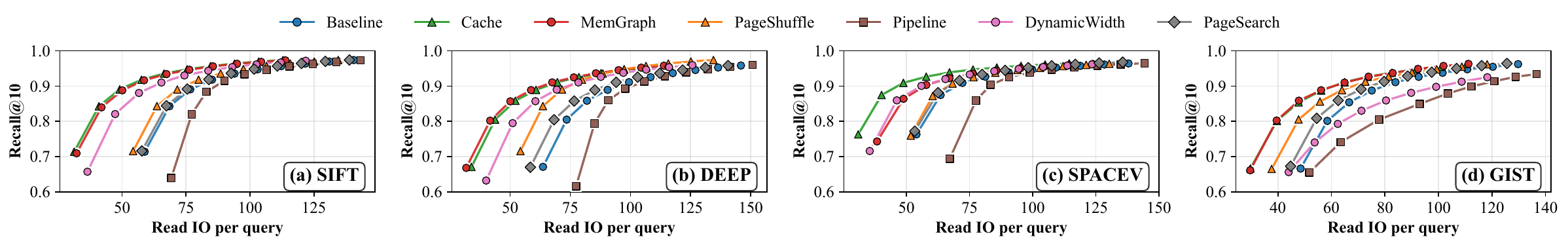}
    \Description{I/O operation count comparison across different optimization algorithms and four datasets, illustrating I/O efficiency improvements.}
    \mycaption{0}{-15}{I/O operations per query across four datasets (Recall@10 matched).}
    \label{fig:io_count_comparison}
\end{figure*}

\section{Individual Optimization Evaluation}
\label{sec:exp-1}

In this section, we evaluate seven representative methods introduced in \S\ref{sec:design_space}: \texttt{Baseline (PQ)}, \texttt{Cache}, \texttt{MemGraph}, \texttt{PageShuffle}, \texttt{DynamicWidth}, \texttt{Pipeline}, and \texttt{PageSearch}.

\subsection{Search Performance Analysis}
On top of the \texttt{PQ} baseline, we ablate each of the six techniques to quantify their standalone effects. For completeness, these techniques span all three dimensions of our taxonomy: memory layout (\texttt{Cache, MemGraph}), disk layout (\texttt{PageShuffle}), and search algorithm optimization (\texttt{Pipeline, DynamicWidth, PageSearch}). We report Pareto fronts across multiple recall targets in Figures~\ref{fig:recall_qps_comparison}, \ref{fig:latency_recall_comparison}, and \ref{fig:io_count_comparison}, covering throughput (QPS), average latency, and I/O per query, respectively. Intuitively, Figures~\ref{fig:latency_recall_comparison} and~\ref{fig:io_count_comparison} exhibit highly consistent trends, reinforcing the latency breakdown in Figure~\ref{fig:io_profiling}. Because end-to-end latency scales with the number of I/O operations, improving I/O yields proportional latency reductions, validating our I/O-centric optimization focus.

\begin{center}
    \fbox{
    \begin{minipage}{0.95\linewidth}
\textbf{Finding 2:} \emph{\textbf{I/O operations dominate query latency across datasets.} Latency--recall and I/O-per-query curves align closely, indicating that improving I/O translates into proportional latency reductions in disk-based ANN systems.}
    \end{minipage}
    }
\end{center}

Next, we group all those algorithmes into three classes: (1) techniques that deliver obvious performance improvements; (2) techniques with negligible or marginal gains; and (3) techniques that are counterproductive when used alone.

\stitle{Effective single-factor optimizations.}
Optimizations targeting memory layout and search-width adaptivity provide the strongest standalone gains across all four datasets. With SIFT, \texttt{MemGraph} reduces I/O by about \textbf{45\%} and increases QPS by about \textbf{90\%} at relaxed accuracy (\eg \(L=10\) with \(\text{Recall@10}\approx 0.7\)); even at high accuracy (\eg \(L=100\) with \(\text{Recall@10}\ge 0.95\)), it still reduces I/O by about \textbf{19\%} and boosts throughput by about \textbf{25\%}. Its overhead is modest: sampling 0.1\% vertices for an in-memory navigation graph adds only \(\sim\)30-50\,MB (cf. Table~\ref{tab:constrction_overhead_comparison}). The \texttt{Cache} policy accelerates early-hop I/O by retaining nodes near entry points in memory. 
Because this SSSP-based caching is sensitive to graph quality, it delivers especially strong gains on SPACEV, as SPACEV's graph index contains many low-degree nodes that concentrate early-hop paths near the entry points.

\begin{figure}[t]
    \centering
    \includegraphics[width=\linewidth]{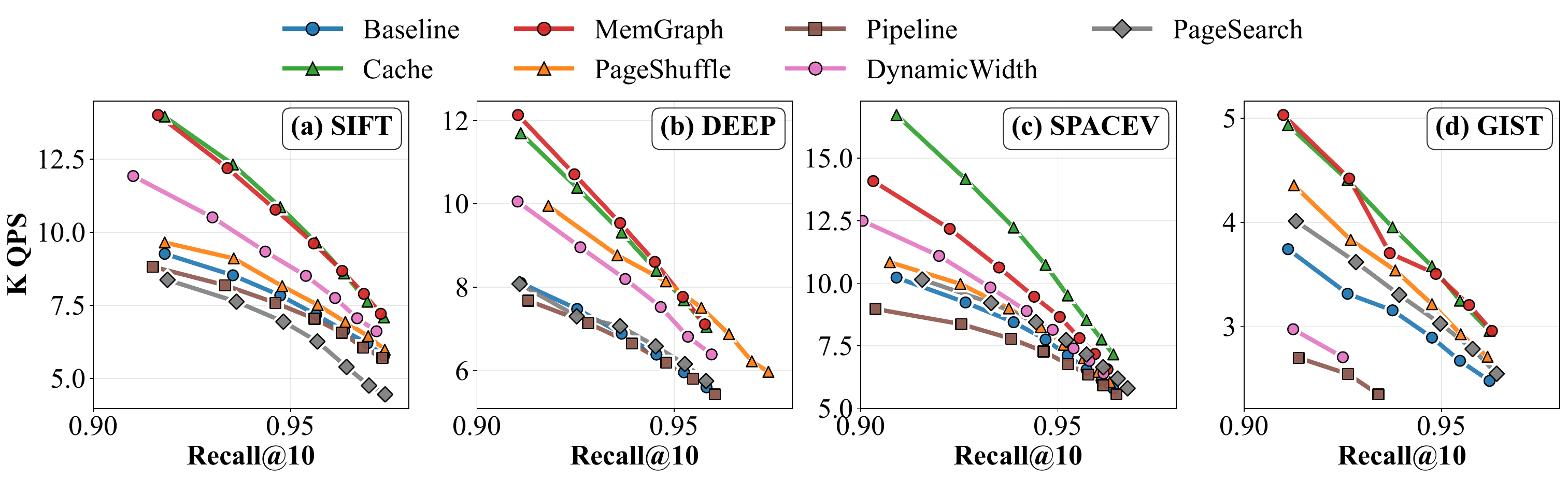}
    \Description{Recall--QPS trade-off across seven optimizations and four datasets (Recall@10 matched).}
    \mycaption{0}{-20}{Zoom view for Recall@10 $\ge$ 0.90 in Figure~\ref{fig:recall_qps_comparison}.}
    \label{fig:recall_qps_comparison_high_recall}
\end{figure}

Finally, \texttt{DynamicWidth} improves I/O utilization by adapting the search width between the approach and converge phases: at relaxed recall requirements, it reduces I/O operations by about \textbf{25\%} and raises QPS by about \textbf{50\%}. This adaptivity can, however, slightly reduce accuracy compared with a fixed width \(\omega\) configuration; \eg on GIST with \(L=10\), recall decreases from \textbf{66\%} to \textbf{50\%}. 
On high-dimensional datasets with high recall requirements, \texttt{DynamicWidth} exhibits severe performance degradation (\cf Fig.~\ref{fig:recall_qps_comparison_high_recall}). This occurs because the number of search iterations increases significantly, causing the search width to grow progressively larger and consequently triggering substantially more I/O operations. Therefore, in such scenarios, we do not recommend using \texttt{DynamicWidth} in isolation; instead, it should be combined with other techniques that reduce hop counts.

\begin{center}
\fbox{
\begin{minipage}{0.95\linewidth}
\textbf{Finding 3:} \emph{\textbf{The strongest single-factor gains come from memory layout and search width adaptivity.} \texttt{MemGraph} and \texttt{Cache} reduce I/O by providing better entry points that shorten disk traversal paths, while \texttt{DynamicWidth} reduces I/O by eliminating wasteful exploration in early search phases. These optimizations target the root causes of I/O inefficiency: poor starting positions and over-expansion during approach phases.}
\end{minipage}}
\end{center}

\stitle{Optimizations with limited standalone impact.}
\texttt{PageShuffle} increases inter page neighbor locality, yet average I/O drops by only \(\sim\)\textbf{9\%} on \textsc{SIFT}. In beam search, most I/O is incurred by pages along the multi-hop expansion path; improving one-hop neighbor locality seldom alters that path and therefore does little to reduce page I/O operations. 

\texttt{PageSearch} leaves I/O counts essentially unchanged in our setup but increases in-page computation by computing all records on each fetched page; 
For uint8 datasets such as \textsc{SIFT} and \textsc{SPACEV}, where each page contains more than 10 vector records, \texttt{PageSearch} introduces substantial per-page computational overhead, which sharply degrades SSD performance (\cf Table~\ref{tab:disk_performance_comparison}). The extra computation prolongs SSD idle time and reduces IOPS (\cf Fig.~\ref{fig:overlap_io_and_compute}).
Prior work (\eg Starling) reports strong gains when \texttt{PageShuffle} and \texttt{PageSearch} are combined; we evaluate this claim in the combination study below.

\begin{center}
\fbox{
\begin{minipage}{0.95\linewidth}
\textbf{Finding 4:} \emph{\textbf{\texttt{PageShuffle} and \texttt{PageSearch} underperform in isolation.} Locality gains and per-page computing do not translate into fewer I/Os; \texttt{PageSearch} may even introduce performance degradation on \textsc{SIFT} and \textsc{SPACEV}. Their value may emerge when used together.}
\end{minipage}
}
\end{center}

\stitle{Counterintuitive degradations from advanced techniques.}
As shown in Table~\ref{tab:disk_performance_comparison}, \texttt{Pipeline} fails to improve disk I/O because baseline algorithms already saturate disk performance under high concurrency. Instead, \texttt{Pipeline} widens the exploration frontier by issuing I/O requests before the best candidate is confirmed, leading to speculative reads. This increases the total number of I/O operations (\cf Figure~\ref{fig:io_count_comparison}), which paradoxically degrades end-to-end performance rather than improving it.

\begin{center}
\fbox{
\begin{minipage}{0.95\linewidth}
\textbf{Finding 5:} \emph{\textbf{\texttt{Pipeline} optimization can be counterproductive.} \texttt{Pipeline} widens the exploration frontier by speculative reading, which introduces more IO operations and reduces the overall performance.}
\end{minipage}}
\end{center}

\begin{table}[t]
    \footnotesize 
    \renewcommand\arraystretch{1.0}
    \mycaption{0}{0}{Disk Metrics Comparison.}
       \label{tab:disk_performance_comparison}
    \setlength{\tabcolsep}{0.6mm}{ 
       \begin{tabular}{|l|cccccccc|}
           \hline
                                            & \multicolumn{2}{c|}{\textbf{SIFT}}                                                                                     & \multicolumn{2}{c|}{\textbf{DEEP}}                                        & \multicolumn{2}{c|}{\textbf{SPACEV}}                                                                                   & \multicolumn{2}{c|}{\textbf{GIST}}                                                                \\ \cline{2-9} 
           \multirow{-2}{*}{\textbf{Method}} & \multicolumn{1}{c|}{\textbf{IOPS}}                        & \multicolumn{1}{c|}{\textbf{BW}}                           & \multicolumn{1}{c|}{\textbf{IOPS}} & \multicolumn{1}{c|}{\textbf{BW}}     & \multicolumn{1}{c|}{\textbf{IOPS}}                        & \multicolumn{1}{c|}{\textbf{BW}}                           & \multicolumn{1}{c|}{\textbf{IOPS}}                        & \textbf{BW}                           \\
                                            & \multicolumn{1}{c|}{\textbf{(K)}}                         & \multicolumn{1}{c|}{\textbf{(MB/s)}}                       & \multicolumn{1}{c|}{\textbf{(K)}}  & \multicolumn{1}{c|}{\textbf{(MB/s)}} & \multicolumn{1}{c|}{\textbf{(K)}}                         & \multicolumn{1}{c|}{\textbf{(MB/s)}}                       & \multicolumn{1}{c|}{\textbf{(K)}}                         & \textbf{(MB/s)}                       \\ \hline
           \textbf{Baseline}                 & \multicolumn{1}{c|}{791}                                  & \multicolumn{1}{c|}{3091}                                  & \multicolumn{1}{c|}{819}           & \multicolumn{1}{c|}{3200}            & \multicolumn{1}{c|}{768}                                  & \multicolumn{1}{c|}{3001}                                  & \multicolumn{1}{c|}{313}                                  & 4889                                  \\
           \textbf{Cache}                    & \multicolumn{1}{c|}{808}                                  & \multicolumn{1}{c|}{3156}                                  & \multicolumn{1}{c|}{813}           & \multicolumn{1}{c|}{3177}            & \multicolumn{1}{c|}{815}                                  & \multicolumn{1}{c|}{3182}                                  & \multicolumn{1}{c|}{322}                                  & 5029                                  \\
           \textbf{MemGraph}                 & \multicolumn{1}{c|}{818}                                  & \multicolumn{1}{c|}{3194}                                  & \multicolumn{1}{c|}{814}           & \multicolumn{1}{c|}{3181}            & \multicolumn{1}{c|}{813}                                  & \multicolumn{1}{c|}{3177}                                  & \multicolumn{1}{c|}{318}                                  & 4973                                  \\
           \textbf{PageShuffle}              & \multicolumn{1}{c|}{803}                                  & \multicolumn{1}{c|}{3136}                                  & \multicolumn{1}{c|}{803}           & \multicolumn{1}{c|}{3136}            & \multicolumn{1}{c|}{741}                                  & \multicolumn{1}{c|}{2895}                                  & \multicolumn{1}{c|}{309}                                  & 4832                                  \\
           \textbf{Pipeline}                 & \multicolumn{1}{c|}{806}                                  & \multicolumn{1}{c|}{3197}                                  & \multicolumn{1}{c|}{808}           & \multicolumn{1}{c|}{3158}            & \multicolumn{1}{c|}{776}                                  & \multicolumn{1}{c|}{3030}                                  & \multicolumn{1}{c|}{323}                                  & 5050                                  \\
           \textbf{DynamicWidth}             & \multicolumn{1}{c|}{796}                                  & \multicolumn{1}{c|}{3109}                                  & \multicolumn{1}{c|}{783}           & \multicolumn{1}{c|}{3060}            & \multicolumn{1}{c|}{765}                                  & \multicolumn{1}{c|}{2987}                                  & \multicolumn{1}{c|}{315}                                  & 4930                                  \\
           \textbf{PageSearch}               & \multicolumn{1}{c|}{\cellcolor[HTML]{F6C1BD}\textbf{690}} & \multicolumn{1}{c|}{\cellcolor[HTML]{F6C1BD}\textbf{2698}} & \multicolumn{1}{c|}{796}           & \multicolumn{1}{c|}{3111}            & \multicolumn{1}{c|}{\cellcolor[HTML]{F6C1BD}\textbf{470}} & \multicolumn{1}{c|}{\cellcolor[HTML]{F6C1BD}\textbf{1834}} & \multicolumn{1}{c|}{315}                                  & 4917                                  \\ \hline
           \end{tabular}
    }
    \vspace{-5mm}
\end{table}

\definecolor{Black}{rgb}{0, 0, 0} 
\definecolor{myred}{rgb}{0.965, 0.757, 0.741} 
\definecolor{myblue}{rgb}{0.765, 0.773, 0.949}

\begin{table*}[t]
    \footnotesize 
    \renewcommand\arraystretch{1.0}
    \mycaption{0}{0}{Index Construction Overhead. Columns: \textbf{BT} = total build time (seconds), \textbf{MaxM} = peak memory consumption during construction (GB), \textbf{Disk} = total on-disk index size (GB), \textbf{Mem} = total in-memory auxiliary structure size (GB). Color coding: \fcolorbox{Black}{myred}{{\color{myred}-}} indicates significant overhead; \fcolorbox{Black}{myblue}{{\color{myblue}-}} indicates modest overhead.}
       \label{tab:constrction_overhead_comparison}
    \setlength{\tabcolsep}{1.2mm}{ 
        \begin{tabular}{|l|cccccccccccccccc|}
            \hline
                                              & \multicolumn{4}{c|}{\textbf{SIFT}}                                                                                                                                                                               & \multicolumn{4}{c|}{\textbf{DEEP}}                                                                                                                                                                                & \multicolumn{4}{c|}{\textbf{SPACEV}}                                                                                                                                                                             & \multicolumn{4}{c|}{\textbf{GIST}}                                                                                                                                                        \\ \cline{2-17} 
            \multirow{-2}{*}{\textbf{Method}} & \multicolumn{1}{c|}{\textbf{BT}}                   & \multicolumn{1}{c|}{\textbf{MaxM}}                 & \multicolumn{1}{c|}{\textbf{Disk}}                 & \multicolumn{1}{c|}{\textbf{Mem}}                 & \multicolumn{1}{c|}{\textbf{BT}}                   & \multicolumn{1}{c|}{\textbf{MaxM}}                 & \multicolumn{1}{c|}{\textbf{Disk}}                 & \multicolumn{1}{c|}{\textbf{Mem}}                  & \multicolumn{1}{c|}{\textbf{BT}}                  & \multicolumn{1}{c|}{\textbf{MaxM}}                  & \multicolumn{1}{c|}{\textbf{Disk}}                 & \multicolumn{1}{c|}{\textbf{Mem}}                 & \multicolumn{1}{c|}{\textbf{BT}}                  & \multicolumn{1}{c|}{\textbf{MaxM}}                 & \multicolumn{1}{c|}{\textbf{Disk}}               & \textbf{Mem}                  \\ \hline
            Baseline                          & \multicolumn{1}{c|}{8814}                          & \multicolumn{1}{c|}{31.9}                         & \multicolumn{1}{c|}{39.06}                         & \multicolumn{1}{c|}{3.81}                         & \multicolumn{1}{c|}{14960}                         & \multicolumn{1}{c|}{41.5}                         & \multicolumn{1}{c|}{65.11}                         & \multicolumn{1}{c|}{3.82}                          & \multicolumn{1}{c|}{9393}                         & \multicolumn{1}{c|}{36.4}                           & \multicolumn{1}{c|}{39.31}                         & \multicolumn{1}{c|}{3.81}                         & \multicolumn{1}{c|}{1135}                         & \multicolumn{1}{c|}{14.25}                         & \multicolumn{1}{c|}{5.1}                         & 0.2                           \\ \hline
            Baseline + MemGraph                          & \multicolumn{1}{c|}{\cellcolor[HTML]{C3C5F2}8817}  & \multicolumn{1}{c|}{31.9}                         & \multicolumn{1}{c|}{39.06}                         & \multicolumn{1}{c|}{\cellcolor[HTML]{C3C5F2}3.84} & \multicolumn{1}{c|}{\cellcolor[HTML]{C3C5F2}14981} & \multicolumn{1}{c|}{41.5}                         & \multicolumn{1}{c|}{65.11}                         & \multicolumn{1}{c|}{\cellcolor[HTML]{C3C5F2}3.87}  & \multicolumn{1}{c|}{\cellcolor[HTML]{C3C5F2}9398} & \multicolumn{1}{c|}{36.4}                           & \multicolumn{1}{c|}{39.31}                         & \multicolumn{1}{c|}{\cellcolor[HTML]{C3C5F2}3.84} & \multicolumn{1}{c|}{1135} & \multicolumn{1}{c|}{14.25}                         & \multicolumn{1}{c|}{5.1}                         & 0.204 \\ \hline
            Baseline + MemGraph + PageShuffle                       & \multicolumn{1}{c|}{\cellcolor[HTML]{F6C1BD}11095} & \multicolumn{1}{c|}{\cellcolor[HTML]{F6C1BD}87.7} & \multicolumn{1}{c|}{\cellcolor[HTML]{FFFFFF}39.06} & \multicolumn{1}{c|}{\cellcolor[HTML]{C3C5F2}4.61} & \multicolumn{1}{c|}{\cellcolor[HTML]{F6C1BD}17303} & \multicolumn{1}{c|}{\cellcolor[HTML]{F6C1BD}90.8} & \multicolumn{1}{c|}{\cellcolor[HTML]{FFFFFF}65.11} & \multicolumn{1}{c|}{\cellcolor[HTML]{C3C5F2}4.70} & \multicolumn{1}{c|}{\cellcolor[HTML]{F6C1BD}11760} & \multicolumn{1}{c|}{\cellcolor[HTML]{F6C1BD}109.66} & \multicolumn{1}{c|}{\cellcolor[HTML]{FFFFFF}39.31} & \multicolumn{1}{c|}{\cellcolor[HTML]{C3C5F2}4.61} & \multicolumn{1}{c|}{\cellcolor[HTML]{F6C1BD}1158} & \multicolumn{1}{c|}{14.25} & \multicolumn{1}{c|}{5.1} & 0.209 \\ \hline
            \end{tabular}
    }
    \vspace{-3mm}
\end{table*}

\subsection{Graph Index Construction Overhead}
As shown in Table~\ref{tab:constrction_overhead_comparison}, we evaluate the index construction overhead, including build time, peak memory consumption during construction, and index structure size (covering both on-disk index storage and in-memory auxiliary structures). The \texttt{Baseline} configuration represents DiskANN's original disk structure construction. Among the evaluated techniques, \texttt{MemGraph} and \texttt{PageShuffle} require additional offline preprocessing, while the other methods are query-time optimizations that incur no extra construction overhead. \texttt{MemGraph} constructs a small in-memory navigation graph by sampling a subset of vertices; its total overhead is modest. However, \texttt{PageShuffle} imposes substantial costs. First, the shuffling phase is time-consuming because it involves multiple iterations to solve an NP-hard optimization problem. Second, its memory footprint is large: the algorithm must load both the full graph and its reverse graph into memory, significantly increasing peak memory requirements—this constraint explains why our evaluation is limited to datasets up to 100 million vectors. Additionally, since records are reordered, the system must maintain an in-memory mapping structure between record IDs and page IDs, which further enlarging the memory index footprint.

\begin{center}
\fbox{
\begin{minipage}{0.95\linewidth}
\textbf{Finding 6:} \emph{\textbf{\texttt{PageShuffle} is both time-intensive and memory-intensive.} If practitioners adopt \texttt{PageShuffle} to mitigate DiskANN's read amplification by reorganizing page layouts, they must allocate additional build time and carefully evaluate memory capacity constraints, particularly for large-scale deployments.}
\end{minipage}
}
\end{center}

\subsection{Memory Budget Analysis}
As we know, increasing memory budgets yields substantial performance improvements: larger \texttt{MemGraph} sampling graphs provide better performance gains, and allocating more memory to \texttt{PQ} enables higher-dimensional quantized vectors, thereby improving search accuracy. However, when memory headroom exists, the question of how to optimally allocate budgets between \texttt{PQ} and \texttt{MemGraph} warrants investigation. 

Quantized coordinates require storing compressed representations for every vector in the dataset, achieving compression ratios of tens-to-one compared to the original data, yet the resulting memory footprint remains substantial. In contrast, \texttt{MemGraph} samples a small fraction of the original graph and stores only topological information (no coordinate data), resulting in memory overhead that is typically much smaller than \texttt{PQ} compressed coordinates. 

\begin{figure}[t]
    \centering
    \includegraphics[width=\linewidth]{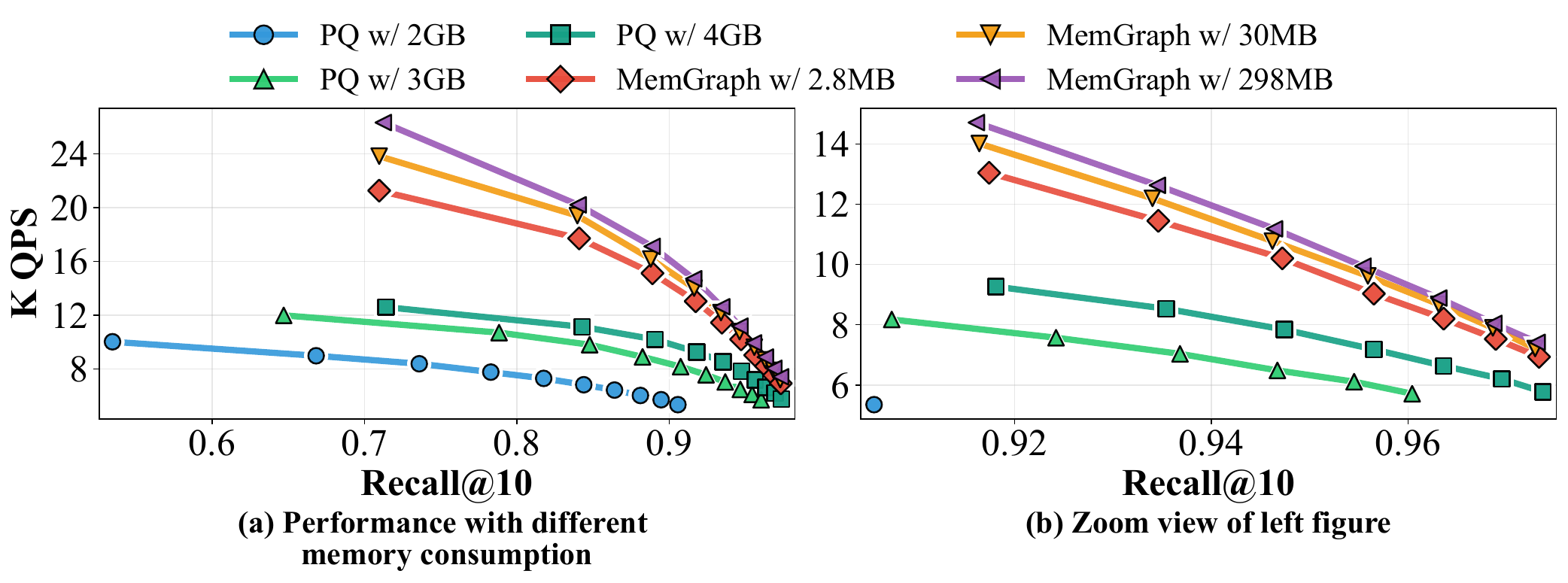}
    \mycaption{0}{-18}{Memory effect on performance.}
    \label{fig:mem_graph_plot}
\end{figure}

Figure~\ref{fig:mem_graph_plot} compares performance under different \texttt{PQ} coordinate and \texttt{MemGraph} memory budgets on the \textsc{SIFT} dataset. Two key observations emerge. First, under identical search conditions, increasing \texttt{PQ} coordinate dimensions improves recall accuracy but has minimal impact on throughput. Second, increasing \texttt{MemGraph} sampling ratios (the figure shows ratios of 1/10{,}000, 1/1{,}000, and 1/100) substantially improves throughput but has little effect on recall. At high sampling ratios, further increasing the sampling frequency yields diminishing performance returns.

\begin{center}
    \fbox{
    \begin{minipage}{0.95\linewidth}
    \textbf{Finding 7:} \emph{\textbf{When memory headroom exists, prioritize allocating a modest budget to \texttt{MemGraph} sampling, then incrementally increase \texttt{PQ} coordinate memory to achieve optimal performance.}}
    \end{minipage}
    }
\end{center}

\begin{figure*}[t]
    \centering
    \includegraphics[width=\linewidth]{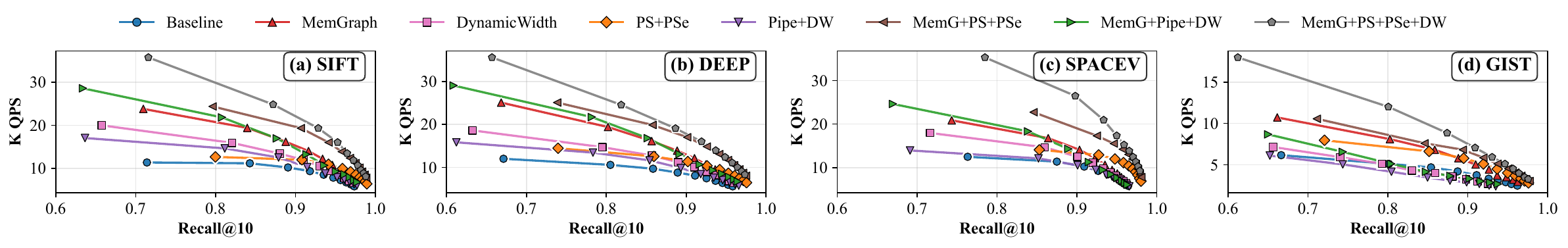}
    \Description{Recall versus QPS comparison across five optimization combinations (C1--C5) and four datasets, showing accuracy--throughput trade-offs.}
    \mycaption{0}{-15}{Recall--QPS trade-off: five combinations (C1--C5) across four datasets.}
    \label{fig:combination_recall_qps_comparison}
\end{figure*}

\begin{figure*}[t]
    \centering
    \includegraphics[width=\linewidth]{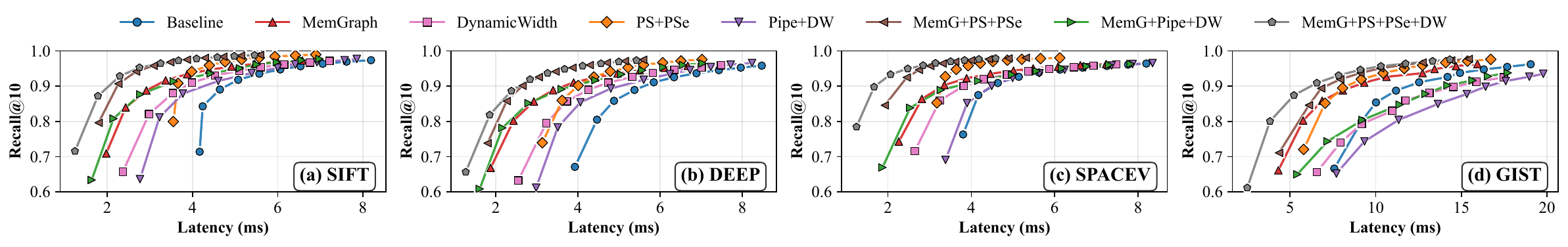}
    \Description{Latency vs. Recall across five optimization combinations (C1--C5) and four datasets, demonstrating performance--accuracy relationships.}
    \mycaption{0}{-15}{Latency--Recall@10 trade-off: five combinations (C1--C5) across four datasets.}
    \label{fig:combination_latency_recall_comparison}
\end{figure*}

\begin{figure}[t]
    \centering
    \includegraphics[width=\linewidth]{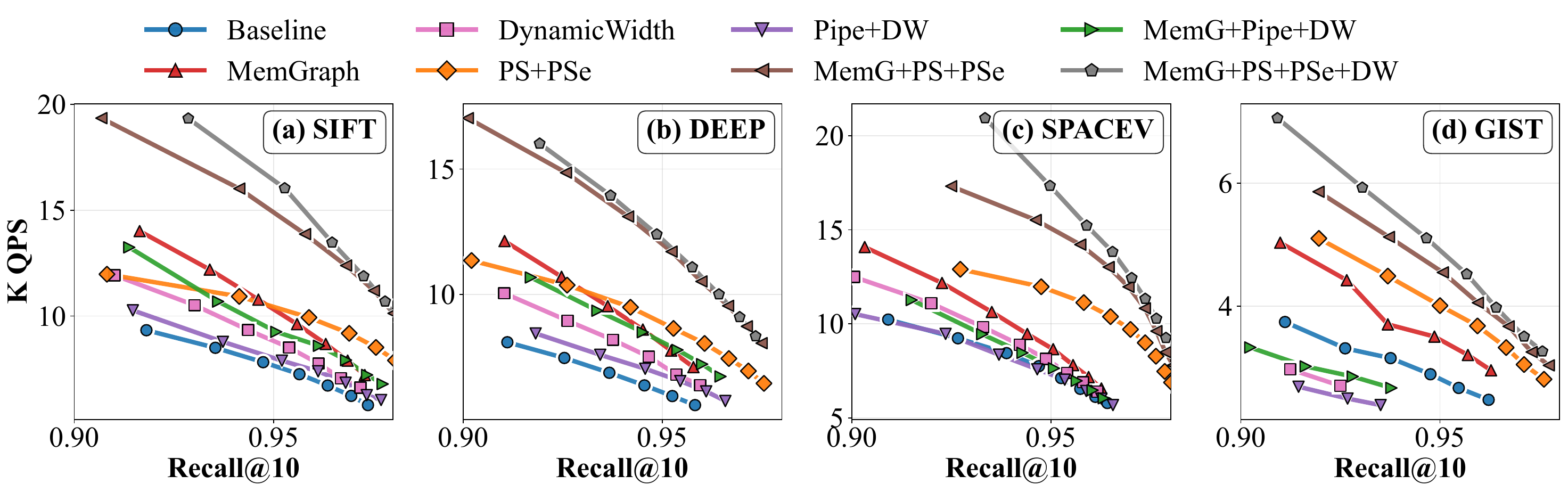}
    \Description{Zoom view for Recall@10 $\ge$ 0.90 in Figure~\ref{fig:combination_recall_qps_comparison}.}
    \mycaption{0}{-15}{Zoom view for Recall@10 $\ge$ 0.90 in Figure~\ref{fig:combination_recall_qps_comparison}.}
    \label{fig:combination_recall_qps_comparison_high_recall}
\end{figure}

\section{Combination Effect Evaluation}
\label{sec:exp-2}

This section evaluates representative multi-factor combinations across memory layout, disk layout, and search scheduling, reports their cumulative effects and interactions, and compares the best combination against state-of-the-art systems.
\subsection{Combination Evaluation}

\subsubsection{Combination Design}
To keep the study tractable and, more importantly, interpretable, we select a small set of combinations guided by three principles: (i) complementarity—prioritizing pairs whose goals are orthogonal and therefore likely to yield synergy, (ii) minimal stepwise deltas—adding at most one new factor at a time so that marginal effects remain attributable.

Concretely, we exclude two techniques from the combination space: \texttt{AiS}, which incurs substantial on-disk expansion and read amplification; and \texttt{SSSP-based Caching}, which is less flexible in our setting due to static entry-point dependency and is rarely adopted in recent systems compared with \texttt{MemGraph} (\cf Table~\ref{tab:algorithm_optimization_comparison}). As references, we keep three single-factor baselines from \S\ref{sec:experiments}: \texttt{Baseline} (PQ), \texttt{MemGraph}, and \texttt{DynamicWidth}.

Based on these criteria, we instantiate five progressive combinations beyond the PQ baseline to reveal cumulative effects and potential conflicts/synergies. Specifically, we consider \textbf{C1} (PS + PSe), which enables \texttt{PageShuffle} and \texttt{PageSearch} on top of \texttt{Baseline}; \textbf{C2} (Pipe + DW), which enables \texttt{Pipeline} and \texttt{DynamicWidth} on top of \texttt{Baseline}; \textbf{C3} (MemG + PS + PSe), which adds \texttt{MemGraph} to C1; \textbf{C4} (MemG + Pipe + DW), which adds \texttt{MemGraph} to C2; and \textbf{C5} (MemG + PS + PSe + DW), which combines \texttt{MemGraph} with PS, PSe, and DynamicWidth.
Pilot sweeps over additional permutations (e.g., PS+DW, PSe+DW, PS+Pipe) did not change conclusions and are omitted for brevity and space constraints.

\begin{table}[t]
    \footnotesize 
    \renewcommand\arraystretch{0.6}
    \caption{{Disk Metrics for five combination algorithms.}}
           \label{tab:combination_resource_utilization}
    \setlength{\tabcolsep}{0.5mm}{ 
       \begin{tabular}{|l|cccccccc|}
           \hline
                                       & \multicolumn{2}{c|}{\textbf{SIFT}}                                                      & \multicolumn{2}{c|}{\textbf{DEEP}}                                        & \multicolumn{2}{c|}{\textbf{SPACEV}}                                                    & \multicolumn{2}{c|}{\textbf{GIST}}                   \\ \cline{2-9} 
                                       & \multicolumn{1}{c|}{\textbf{IOPS}}               & \multicolumn{1}{c|}{\textbf{BW}}     & \multicolumn{1}{c|}{\textbf{IOPS}} & \multicolumn{1}{c|}{\textbf{BW}}     & \multicolumn{1}{c|}{\textbf{IOPS}}               & \multicolumn{1}{c|}{\textbf{BW}}     & \multicolumn{1}{c|}{\textbf{IOPS}} & \textbf{BW}     \\
           \multirow{-3}{*}{\textbf{Method}}    & \multicolumn{1}{c|}{\textbf{(K)}}                & \multicolumn{1}{c|}{\textbf{(MB/s)}} & \multicolumn{1}{c|}{\textbf{(K)}}  & \multicolumn{1}{c|}{\textbf{(MB/s)}} & \multicolumn{1}{c|}{\textbf{(K)}}                & \multicolumn{1}{c|}{\textbf{(MB/s)}} & \multicolumn{1}{c|}{\textbf{(K)}}  & \textbf{(MB/s)} \\ \hline
           \textbf{Baseline}           & \multicolumn{1}{c|}{791}                         & \multicolumn{1}{c|}{3091}            & \multicolumn{1}{c|}{819}           & \multicolumn{1}{c|}{3200}            & \multicolumn{1}{c|}{768}                         & \multicolumn{1}{c|}{3001}            & \multicolumn{1}{c|}{313}           & 4889            \\
           \textbf{MemGraph}           & \multicolumn{1}{c|}{818}                         & \multicolumn{1}{c|}{3194}            & \multicolumn{1}{c|}{814}           & \multicolumn{1}{c|}{3181}            & \multicolumn{1}{c|}{813}                         & \multicolumn{1}{c|}{3177}            & \multicolumn{1}{c|}{318}           & 4973            \\
           \textbf{DynamicWidth}       & \multicolumn{1}{c|}{796}                         & \multicolumn{1}{c|}{3109}            & \multicolumn{1}{c|}{783}           & \multicolumn{1}{c|}{3063}            & \multicolumn{1}{c|}{765}                         & \multicolumn{1}{c|}{2987}            & \multicolumn{1}{c|}{315}           & 4930            \\
           \textbf{PS+PSe}         & \multicolumn{1}{c|}{\cellcolor[HTML]{FFCCC9}756} & \multicolumn{1}{c|}{2954}            & \multicolumn{1}{c|}{809}           & \multicolumn{1}{c|}{3160}            & \multicolumn{1}{c|}{\cellcolor[HTML]{FFCCC9}705} & \multicolumn{1}{c|}{2755}            & \multicolumn{1}{c|}{309}           & 4828            \\
           \textbf{Pipe+DW}        & \multicolumn{1}{c|}{829}                         & \multicolumn{1}{c|}{3238}            & \multicolumn{1}{c|}{827}           & \multicolumn{1}{c|}{3230}            & \multicolumn{1}{c|}{811}                         & \multicolumn{1}{c|}{3166}            & \multicolumn{1}{c|}{321}           & 5020            \\
           \textbf{MemG+PS+PSe}    & \multicolumn{1}{c|}{\cellcolor[HTML]{FFCCC9}751} & \multicolumn{1}{c|}{2934}            & \multicolumn{1}{c|}{793}           & \multicolumn{1}{c|}{3099}            & \multicolumn{1}{c|}{\cellcolor[HTML]{FFCCC9}683} & \multicolumn{1}{c|}{2668}            & \multicolumn{1}{c|}{312}           & 4872            \\
           \textbf{MemG+Pipe+DW}   & \multicolumn{1}{c|}{818}                         & \multicolumn{1}{c|}{3202}            & \multicolumn{1}{c|}{816}           & \multicolumn{1}{c|}{3189}            & \multicolumn{1}{c|}{826}                         & \multicolumn{1}{c|}{3226}            & \multicolumn{1}{c|}{311}           & 4951            \\
           \textbf{MemG+PS+PSe+DW} & \multicolumn{1}{c|}{\cellcolor[HTML]{FFCCC9}704} & \multicolumn{1}{c|}{2748}            & \multicolumn{1}{c|}{745}           & \multicolumn{1}{c|}{2910}            & \multicolumn{1}{c|}{\cellcolor[HTML]{FFCCC9}643} & \multicolumn{1}{c|}{2512}            & \multicolumn{1}{c|}{314}           & 4909            \\ \hline
           \end{tabular}
    }
\end{table} 

\subsubsection{Combination Optimization Performance Analysis}
By examining Figures~\ref{fig:combination_recall_qps_comparison}--\ref{fig:combination_latency_recall_comparison} and Table~\ref{tab:combination_resource_utilization}, we obtain three main observations:

\stitle{PageShuffle + PageSearch exhibit strong complementarity.}
\texttt{PS} increases the likelihood that useful candidates are colocated within fetched pages; \texttt{PSe} then fully exploits each page by computing all in-page records. As shown in Figures~\ref{fig:combination_recall_qps_comparison} and~\ref{fig:combination_latency_recall_comparison}, under the same $L$ constraint, \textbf{C1} attains higher recall (\eg about a 10\% gain on \textsc{SIFT}); moreover, even at high-accuracy settings (\eg $L=100$ with $\text{Recall}\approx 98\%$), \textbf{C1} still delivers roughly a \textbf{27}\% QPS improvement. Adding \texttt{MemGraph} (\aka \textbf{C3}) further reduces the number of required hops by improving entry-point quality, compounding the benefit.

\begin{center}
\fbox{
\begin{minipage}{0.95\linewidth}
\textbf{Finding 8:} \emph{\textbf{PS + PSe deliver robust synergy: PS raises locality while PSe raises per-page utility.} Their goals are complementary, jointly mitigating locality waste and underutilized pages. With \texttt{MemGraph} (C3), both throughput and latency improve further (vs. PQ baseline at matched Recall@10).}
\end{minipage}
}
\end{center}

 \stitle{Pipeline and DynamicWidth are complementary but not coequal.} \texttt{DynamicWidth} substantially mitigates the useless page reads induced by speculative I/O in pipeline search, making it a natural complement to Pipeline. In the single‑factor results, Pipeline alone underperforms the \texttt{PQ} baseline (see Figure~\ref{fig:recall_qps_comparison}); in the combination study, however, \textbf{C2} improves over the baseline by roughly \textbf{38\%} at matched accuracy (Figure~\ref{fig:combination_recall_qps_comparison}). Nevertheless, \textbf{C2} still trails the DynamicWidth‑only configuration, which is consistent with Finding~5: under common high‑concurrency conditions, \texttt{Pipeline} is often counterproductive and should not be enabled.
For high-dimensional datasets, practitioners should tune the growth rate of \texttt{DynamicWidth} to mitigate potential adverse effects.

\begin{center}
    \fbox{
    \begin{minipage}{0.95\linewidth}
    \textbf{Finding 9:} \emph{\textbf{Pipeline and DynamicWidth are complementary, but DynamicWidth dominates.} C2 (Pipe + DW) improves over the PQ baseline yet remains below DynamicWidth-only. Due to speculative reads and contention, Pipeline is unsuitable under concurrency; enable it only for single-thread, device-exclusive cases.}
    \end{minipage}
    }
    \end{center}
    
    \stitle{Full combination (C5) and definition of OctopusANN.} Integrating \texttt{MemGraph} with \texttt{PageShuffle}, \texttt{PageSearch}, and \texttt{DynamicWidth} achieves the best overall results among the evaluated combinations (throughput, latency, and I/O performance) under our settings, while keeping resource overhead modest (see Figures~\ref{fig:combination_recall_qps_comparison}--\ref{fig:combination_latency_recall_comparison} and Table~\ref{tab:combination_resource_utilization}). The gains arise from complementary effects—higher page-level utilization (PS + PSe) and shorter convergence paths (\texttt{MemGraph} + \texttt{DynamicWidth})—that together amplify I/O effectiveness. We refer to this full combination (C5) as \textbf{OctopusANN}. On \textsc{SPACEV}, C5 attains up to a \textbf{133\%} improvement (Figure~\ref{fig:combination_recall_qps_comparison_high_recall}). The additional cost is minor: small memory and disk overheads, plus extra offline building time for page shuffle during index construction.
    
    \begin{center}
    \fbox{
    \begin{minipage}{0.95\linewidth}
    \textbf{Finding 10:} \emph{The multi-positive combination PQ + MemGraph + PS + PSe + DynamicWidth (C5) delivers the strongest overall results with modest cost—small memory/disk overheads and extra offline page shuffle time—while the online benefits dominate via better page utilization and faster convergence (vs. PQ baseline at matched Recall@10).}
    \end{minipage}
    }
    \end{center}

\subsection{State-of-the-Art Comparison}
\begin{figure}[t]
    \centering
    \includegraphics[width=\linewidth]{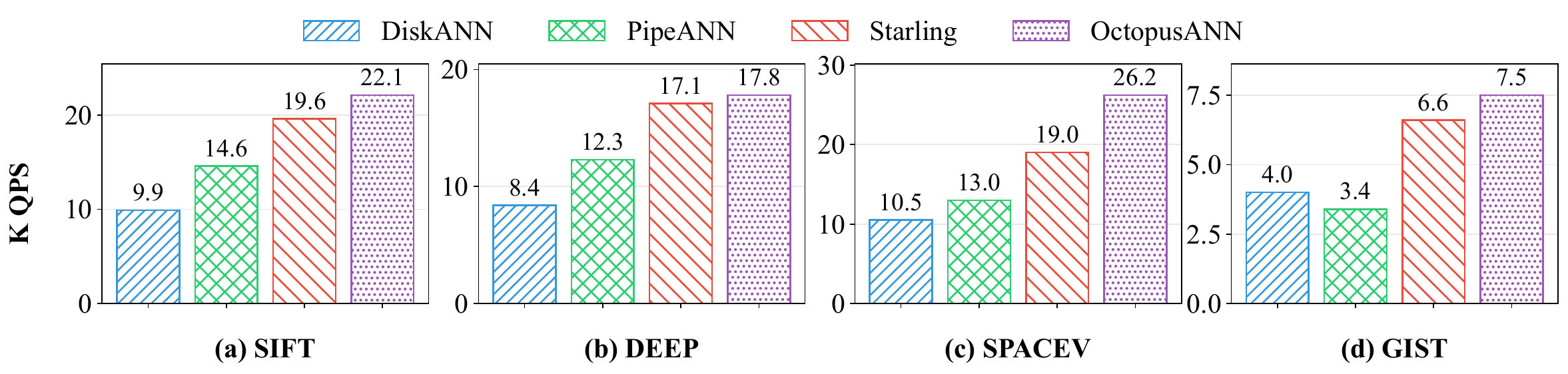}
    \Description{QPS comparison of state-of-the-art algorithms at 90\% recall across four datasets; OctopusANN shows higher throughput.}
    \mycaption{0}{-15}{QPS comparison of state-of-the-art algorithms at 90\% recall across 4 datasets; OctopusANN vs. Starling (+16.8--32.3\%) and vs. DiskANN (+34.4--86.6\%).}
    \label{fig:sota_qps_comparison_90_recall}
\end{figure}

\begin{figure}[t]
    \centering
    \includegraphics[width=\linewidth]{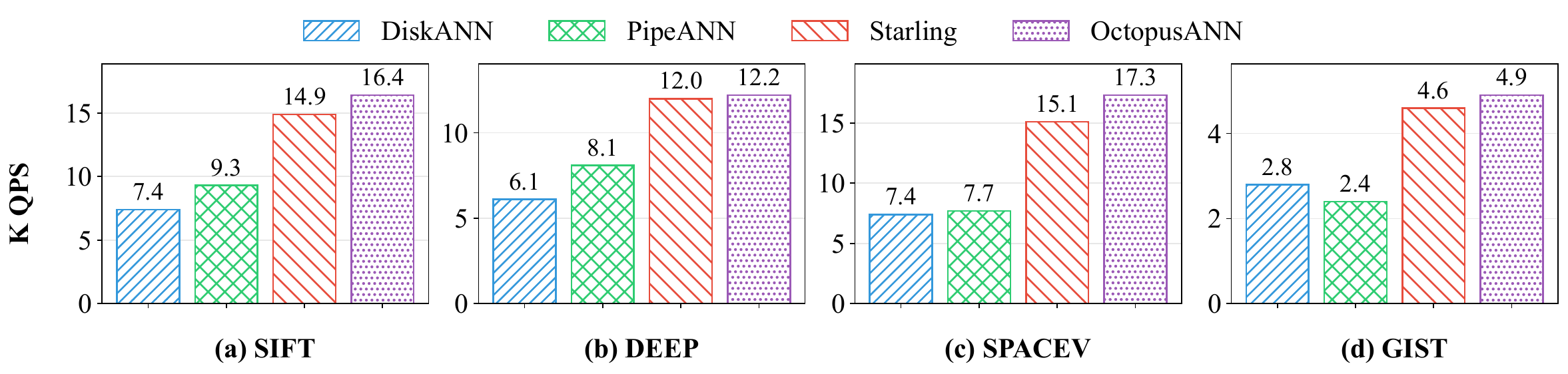}
    \Description{QPS comparison of state-of-the-art algorithms at 95\% recall across four datasets; OctopusANN shows higher throughput.}
    \mycaption{0}{-15}{QPS comparison of state-of-the-art algorithms at 95\% recall across four datasets; OctopusANN shows higher throughput.}
    \label{fig:sota_qps_comparison_95_recall}
\end{figure}

We compare our best combination (OctopusANN, C5) against state-of-the-art disk-based ANN systems at \textbf{Recall@10=90\%} and \textbf{Recall@10=95\%}. The competitors include \texttt{DiskANN}~\cite{Subramanya2019}, \texttt{Starling}~\cite{Wang2024}, and \texttt{PipeANN}~\cite{guoachieving}. At \textbf{Recall@10=90\%}, OctopusANN delivers \textbf{4.1--37.9\%} higher QPS than \texttt{Starling}, and \textbf{87.5--149.5\%} over \texttt{DiskANN} (Figure~\ref{fig:sota_qps_comparison_90_recall}). At the more stringent \textbf{Recall@10=95\%}, the throughput advantage over \texttt{Starling} narrows to \textbf{1.6--14.6\%} (Figure~\ref{fig:sota_qps_comparison_95_recall}), indicating diminishing gains for \texttt{DynamicWidth} at higher accuracy.

\begin{figure}[t]
    \centering
    \includegraphics[width=\linewidth]{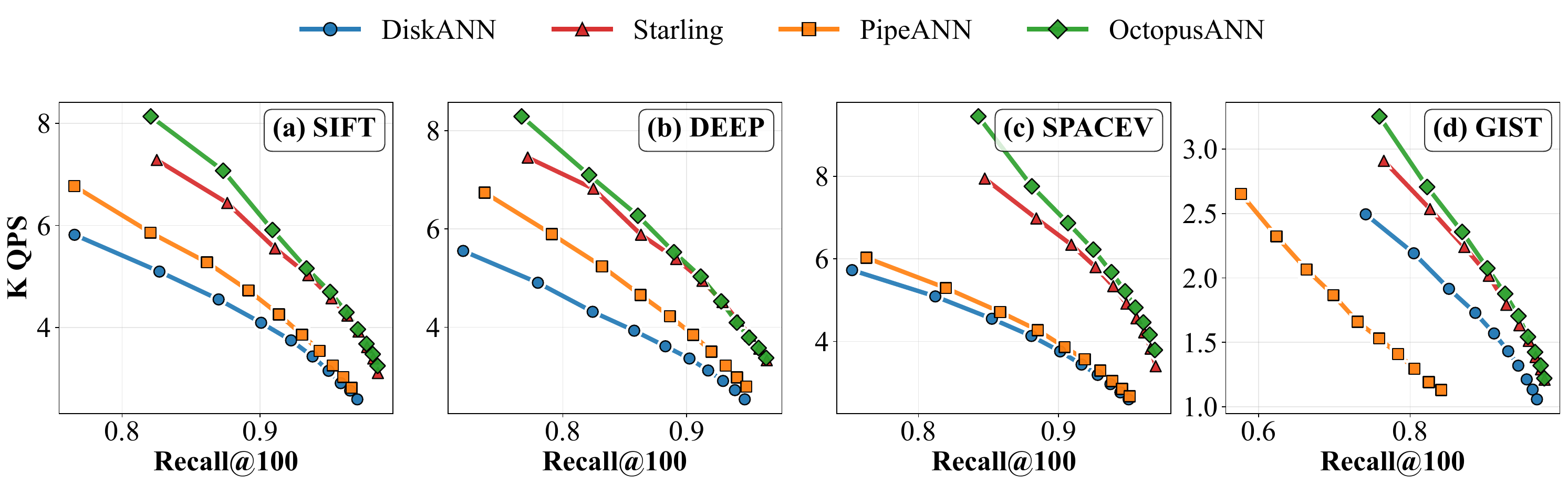}
    \Description{QPS comparison of top 100 across 4 systems.}
    \mycaption{0}{-10}{QPS comparison of top 100 across 4 systems.}
    \label{fig:top100_recall_qps_comparison}
\end{figure}

Figure~\ref{fig:top100_recall_qps_comparison} shows the QPS comparison of four algorithms on the top-100k dataset. The relative performance trends among algorithms are similar to those observed for top-10 queries. However, for high-recall requirements (\eg Recall@100 > 90\%), OctopusANN shows less pronounced improvements. This is because the effectiveness of \texttt{DynamicWidth} diminishes as the number of search iterations increases, and top-100 queries typically require substantially more iterations than top-10 queries.

\begin{figure}[t]
    \centering
    \includegraphics[width=\linewidth]{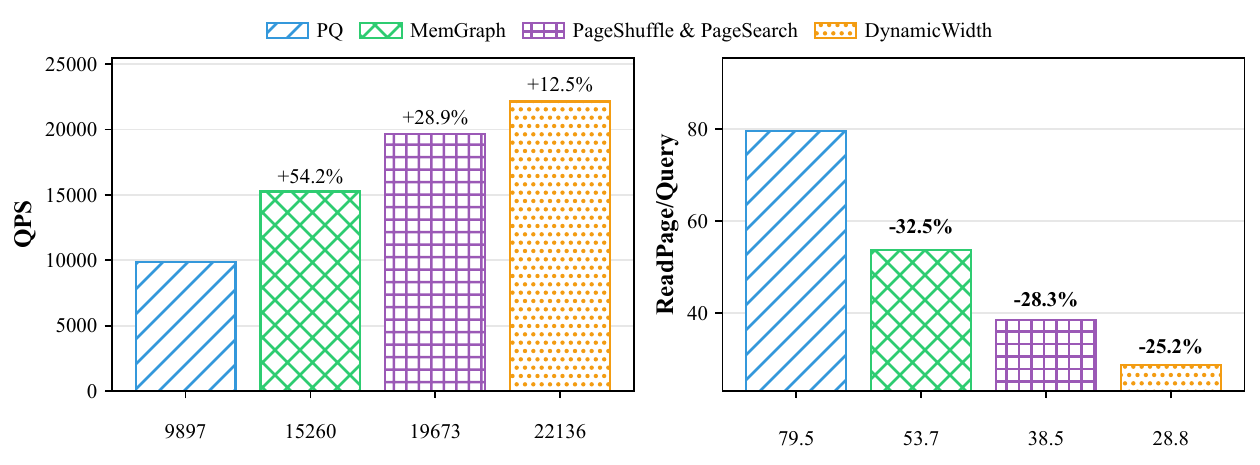}
    \Description{Breakdown of the optimization combination: incremental QPS improvements and reduction in read pages per query.}
    \mycaption{0}{-10}{OctopusANN optimization breakdown analysis on SIFT dataset.}
    \label{fig:c11_breakdown_analysis}
\end{figure}

\stitle{OctopusANN breakdown analysis.}
To clarify the cumulative effects of our systematic optimizations, Figure~\ref{fig:c11_breakdown_analysis} presents a breakdown of the \texttt{OctopusANN} strategy and quantifies each technique's incremental contribution to overall performance. The leftmost bar denotes the \texttt{baseline}; subsequent bars add optimizations cumulatively from left to right.

The breakdown analysis reveals critical insights into our systematic optimization approach. The QPS improvements (left) show that \texttt{MemGraph} provides the foundation with \textbf{54.2\%} improvement over \texttt{baseline}, followed by \texttt{PageShuffle\&PageSearch} (\textbf{28.9\%}), and \texttt{DynamicWidth} (\textbf{12.5\%}). The I/O performance analysis (right) demonstrates dramatic reductions in read pages per query: \texttt{MemGraph} reduces pages by \textbf{32.5\%}, \texttt{PageShuffle\&PageSearch} by \textbf{28.3\%}, and \texttt{DynamicWidth} by \textbf{25.2\%}. This dual perspective validates that performance gains stem from I/O optimization.

\begin{center}
    \fbox{
    \begin{minipage}{0.95\linewidth}
    \textbf{Finding 11:} \emph{\textbf{Systematic multi-dimensional optimization is most effective at moderate accuracy}: at 90\% recall, OctopusANN outperforms \texttt{Starling} by \textbf{4.1--37.9\%} and \texttt{DiskANN} by \textbf{87.5--149.5\%}, whereas at 95\% recall the gains are limited.}
    \end{minipage}
    }
\end{center}

\subsection{Discussion}
\begin{figure}[t]
    \centering
    \includegraphics[width=\linewidth]{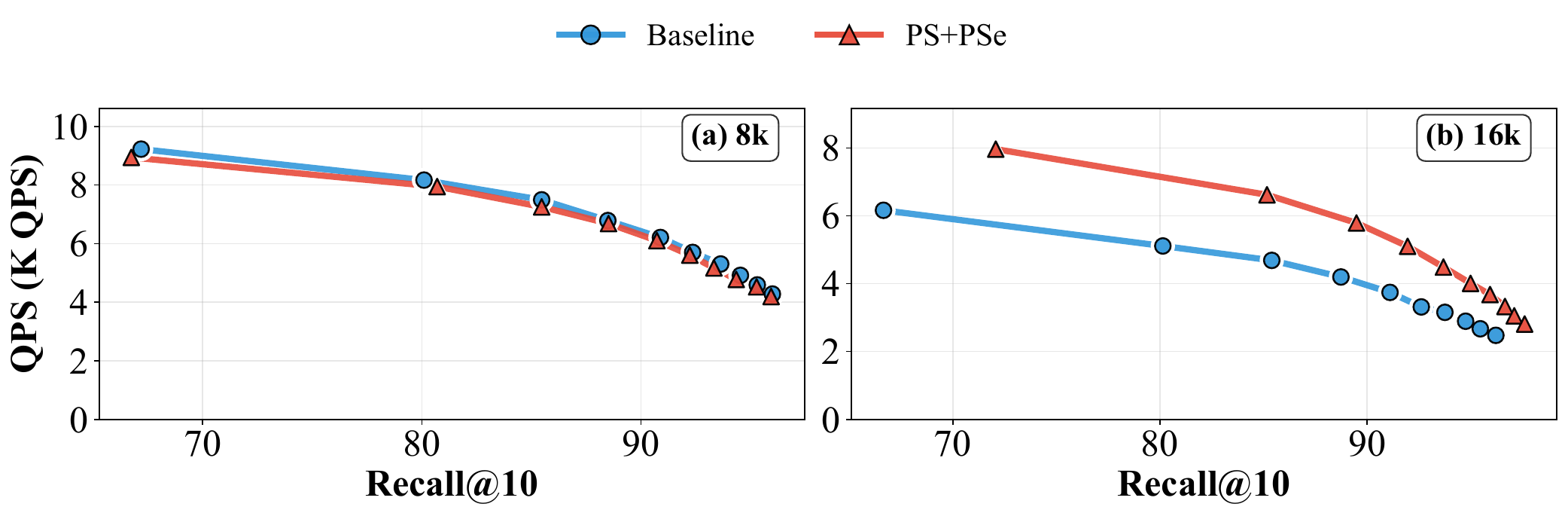}
    \Description{Comparison of four algorithms on GIST8K dataset with different page sizes.}
    \mycaption{0}{-10}{Comparison of baseline and PS+PSe on GIST dataset with different page sizes.}
    \label{fig:gist8k_comparison}
\end{figure}

\stitle{High-Dimensional Data and Page Size Trade-offs.}
For high-dimensional data (\eg approaching or exceeding 1{,}000 dimensions), the standard 4\,KB page size is insufficient to store a single vector record, necessitating larger page sizes. Figure~\ref{fig:gist8k_comparison} compares the QPS of \texttt{Baseline} and \texttt{PS+PSe} on the GIST dataset under 8\,KB and 16\,KB page configurations, revealing three key trade-offs. First, optimization effectiveness varies significantly: with 8\,KB pages, \texttt{PS+PSe} provides no performance improvement because each page accommodates only a single record, rendering intra-page shuffling ineffective; conversely, 16\,KB pages enable effective shuffling by holding multiple records per page. Second, index sizes differ substantially—\textbf{7.7\,GB} for 8\,KB pages versus \textbf{5.1\,GB} for 16\,KB pages—because the smaller page size incurs substantial storage waste from unused space, while 16\,KB pages accommodate three records per page, reducing overhead. Third, despite the storage inefficiency, the 8\,KB configuration achieves better overall query performance than 16\,KB because smaller pages reduce per-I/O latency, even though they increase the total number of I/O operations.

\begin{center}
    \fbox{
    \begin{minipage}{0.95\linewidth}
    \textbf{Finding 12:} \emph{\textbf{For high-dimensional datasets, layout optimizations become less effective.} When each page holds fewer vector records, \texttt{PageShuffle+PageSearch} may not provide significant performance improvements. The I/O amplification problem for high-dimensional data warrants further investigation by the research community.}
    \end{minipage}
    }
\end{center}

\section{Conclusion}\label{sec:conclusion}

\begin{figure}[t]
    \centering
    \includegraphics[width=\linewidth]{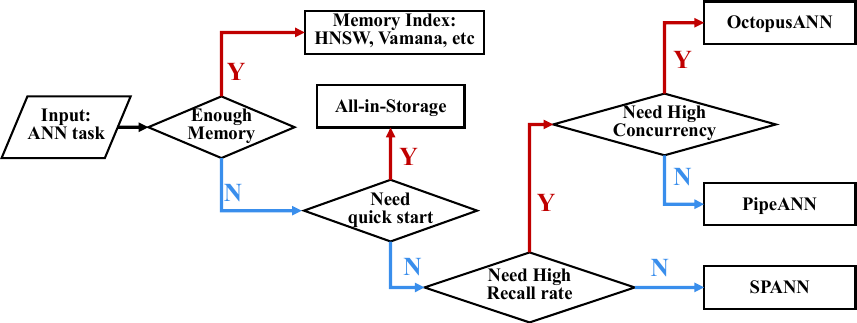}
    \Description{Decision flow for choosing ANN systems: check memory sufficiency; if yes, prefer in-memory indexes (HNSW/Vamana). If memory is tight and quick start is needed, choose All-in-Storage. Otherwise, branch by concurrency and recall requirements: under high concurrency, pick OctopusANN for high recall or MemGraph for moderate recall; under low concurrency, PipeANN is competitive.}
    \mycaption{0}{0}{A practical decision guide for disk-based ANN.}
    \label{fig:guideline}
\end{figure}

In this paper, we present an I/O-first, three-dimensional taxonomy for disk-based ANN and show on 100M-scale datasets that composing techniques across memory layout, disk layout, and search enhances both QPS and end-to-end latency. To translate the findings into practice, Figure~\ref{fig:guideline} provides a concise decision guide. The workflow proceeds as follows: 
(i) when memory is sufficient, prefer mature in-memory indexes (\eg \texttt{HNSW}, \texttt{Vamana}); 
(ii) when memory is tight and fast cold-start is required, \texttt{All-in-Storage} is a convenient choice despite higher on-disk expansion; 
(iii) for low-dimensional datasets under relaxed accuracy targets, inverted-index approaches such as \texttt{SPANN} can be a pragmatic choice; 
(iv) otherwise, decide by concurrency—under low concurrency, \texttt{PipeANN} can be competitive; under high concurrency, choose \texttt{OctopusANN} for high Recall@10 or deploy a lightweight \texttt{MemGraph} when moderate Recall@10 suffices. This matches our empirical results and clarifies when each method is the best fit.

\begin{acks}
This work is supported in part by the National Natural Science Foundation of China under Grant No. U25B2020.
\end{acks}


\balance

\bibliographystyle{ACM-Reference-Format}
\bibliography{main}

\end{document}